\documentclass{ametsocV6.1}

\usepackage{physics}
\usepackage{cancel}
\newcommand{\rhobar}{\overline{\rho}}
\newcommand{\rhobarp}{\overline{\rho}^1}
\newcommand{\rhobarpp}{\overline{\rho}^2}
\newcommand{\irhobar}{\frac{1}{\overline{\rho}}}
\newcommand{\irhobarp}{\frac{1}{\overline{\rho}^1}}
\newcommand{\irhobarpp}{\frac{1}{\overline{\rho}^2}}

\title{Multi-scale dynamics of deep convective self-aggregation. Part II: scale-by-scale budget and mesoscale self-similarity}

\authors{Lorenzo Silvestri,\aff{a}\correspondingauthor{Lorenzo Silvestri, lorenzo.silvestri@unipg.it} 
Paolina Bongioannini Cerlini,\aff{b} 
Andrea Cimarelli\aff{c} 
}

\affiliation{\aff{a}{Department of Civil and Environmental Engineering, University of Perugia}\\
\aff{b}{Department of Physics and Geology, University of Perugia}\\
\aff{c}{Department of Engineering "Enzo Ferrari", University of Modena and Reggio Emilia}
}

\abstract{
This work investigates deep convective self-aggregation in the space of scales, extending the physical-space analysis of Part I. We introduce a novel theoretical framework by examining and deriving a generalized Yaglom equation for the second-order structure function $\langle \delta h_f^{\prime 2} \rangle$, the scale-dependent variance of Frozen Moist Static Energy (FMSE). Self-aggregation is found to be sustained by an inverse energy cascade driven by boundary-layer production and mediated by mesoscale circulations with horizontal scales $r \sim$(100–250 km). The inverse cascade of energy is found to be realized through a series of shallow and mid-level mesoscale circulations which behave in a self-similar way. Accordingly, the scale-energy variance exhibits an extraordinary collapse across all atmospheric layers onto a robust self-similar power law, $\langle \delta h_f^{\prime 2} \rangle \sim r^{6/5}$. In contrast, horizontal velocity increments show a degree of self-similarity ($\langle \delta u_r^{\prime 2} \rangle \sim r^{2/5}$) that retains a residual height dependence. Interestingly, the FMSE scaling breaks down at elevations lower than 750 m and at small local scales below 100 km, allowing us to identify the boundary layer top as a scale-dependent dividing layer where the strong stratification filters the interactions between the surface and free-troposphere, so that small scale dynamics of the low and high atmospheric layers are decoupled while those at large mesoscale are coupled. The derived scaling laws are also used to quantify the moisture-convection feedback via scale-dependent correlation coefficients that show that vertical transport strengthens with scale ($\rho_{mc\perp} \sim r^{2/5}$), while horizontal convergence weakens at the exact inverse rate ($\rho_{mc\parallel} \sim r^{-2/5}$). Framed within convective quasi-equilibrium, these self-similar circulations are found to define a natural dynamical unit of self-aggregation with an upper bound of $\sim 250$ km, suggesting that larger tropical convergence zones emerge from the upscale merging of multiple fundamental units. Overall, the findings enabled by the multi-scale theoretical framework developed here provide a strong theoretical foundation for deep convection and for representing inverse moist energy cascades as key mechanisms for self-aggregation in climate model parameterizations.
} 

\begin{document}

\maketitle

%
%
%
\statement
How atmospheric energy flows across different spatial scales dictates how small cloud systems organize into large clusters. Building on our physical-space insights, this study tracks energy transfer across scales using a generalized fluid-dynamics framework. We discover that self-aggregation is driven by an inverse energy cascade, where energy flows from smaller local clouds to larger mesoscale circulations following a universal scaling law that holds across all atmospheric layers for horizontal scales larger than 100 km. These self-similar circulations define a fundamental "building block" of about 250 km, explaining the natural size limit of tropical cloud clusters. This scale-by-scale framework bridges local cloud physics with planetary-scale climate patterns, offering a powerful foundation for improving how global climate models represent convection.


%
%

%

\section{Introduction}
In idealized numerical simulations mimicking convection over tropical oceans, clouds are found to transition from a random state homogeneously distributed across the domain into a more organized state with a moist convective patch surrounded by a dry patch free of convection. This phenomenon has been termed self-aggregation since such transition occurs only through internal feedbacks related to the convection itself without imposed external large-scale forcings and lateral periodic boundary conditions. Although the occurrence of self-aggregation is robust across a wide range of models \citep{wing2020clouds}, it is very sensitive to domain size, horizontal and vertical grid resolution \citep{muller2012detailed,Yanase2020critical,jenney2023vertical,silvestri2024numerical}, and physical parameterization \citep{shi2021modulation,casallas2025sensitivity}. One relevant feature is the absence of self-aggregation for fine horizontal grid resolutions (less than 1-2 km) and domains smaller than 100 km x 100 km, as shown in the regime diagrams of \citet{muller2012detailed} and \citet{Yanase2020critical} in two different cloud resolving models. This dependence on domain size and resolution is linked to the representation of convective structures that interact at different scales during the simulation: cold pools \citep{jeevanjee2013convective,cerlini2023competing}, convective cores \citep{tompkins2017organization,silvestri2024numerical,casallas2025sensitivity}, circulations \citep{bretherton2005energy,muller2012detailed,Muller2015whatfavors,Yanase2020critical}, dry plumes \citep{yang2020initiation} and low-level clouds \citep{coppin2015physical}. Apart from such sensitivity, at small domain size smaller than 100 km, most models show a natural radiative convective equilibrium state characterized by random convection over the domain. The mean state of such small simulations is usually taken as an initial condition to the large simulations, determining the large-scale condition imposed at the beginning of the simulation. In large domain simulations, the initial random convection aggregates into a single cluster within time-scales of the order of 10 days. Also the mean state of temperature and humidity changes accordingly to the spatial arrangement of convection reaching a final state which has a warmer and drier free troposphere than the beginning. This final state may be regarded as a quasi-steady state, since it is still subjected to oscillations \citep{patrizio2019sensitivity,windmiller2019convection}, but the spatial organization of convection has been proven to be a stable configuration in radiative convective equilibrium (RCE) that is capable of maintaining itself without imposed external forcings. 

This aggregated configuration produces extreme and almost stationary precipitation rates. This, together with its impact on the mean state of the domain, makes such idealized studies relevant to understanding the response of cloud aggregation to global warming \citep{bony2020observed,seidel2020lightness,bao2024intensification}. However, the relevance of such a phenomenon is still debated due to the simplicity of the model set-up with respect to real world realizations \citep{muller2022spontaneous}. The RCE setup has been found to be potentially useful for real convective aggregation at different scales, starting from tropical cyclones \citep{bretherton2005energy,wing2016role} to the Madden-Julian Oscillation \citep[MJO,][]{arnold2015global} and the Inter-Tropical Convergence Zone \citep[ITCZ,][]{muller2020self}. As also underlined in part I of this work \citep{silvestri2026gye_part1}, the ITCZ over tropical oceans is an ideal real world scenario to observe self-aggregation processes. From the large scale point of view, the ITCZ may be viewed as the result of large-scale pattern of Sea Surface Temperature (SST) which forces a low-level inflow of air from the subtropics to the tropics and maintaining convection there \citep{back2009relationship,muller2020self}. However, a closer look at the smaller scales within the ITCZ region can reveal a rich inner life even within its width of about 400 km \citep{Windmiller2024innerlife}, where different convective structures (cold pools and clouds) may combine into a large envelope of convection \citep{windmiller2019convection}. This spontaneous aggregation of smaller structures may occur even in the absence of external forcing \citep{Janssens2024shallow, Bretherton2017mesoscaleagg}. The idealized stationary RCE state considered here, in which a single convective cluster is embedded within dry air, provides a useful analog for the ITCZ as a self-organized tropical regime whose coherence depends on the balance between moisture supply, circulation, and large-scale instabilities. Recent studies have highlighted that the ITCZ can be interpreted through the lens of moisture modes, equatorial Rossby waves, and moisture-vortex instability, all of which can promote disaggregation or reorganize the precipitation structure of the tropical convergence zone \citep{tompkins2025drivers,Adames2024stirring,Maithel2024aggcycles}.

In this context, our focus is on the physical processes that help maintain aggregation during the stationary aggregated phase, rather than on the mechanisms that trigger breakdown. Within this view, the inner convective organization and large-scale maintenance of ITCZ must be intimately connected across scales. Indeed, when the ITCZ breaks down there must exist some transport of energy from the small scale to the large scale of the ITCZ cluster that reestablish its natural aggregated shape. As also demonstrated in the near-global cloud resolving aquaplanet simulation by \cite{bretherton2015convective}, such transport cannot be directly related to radiative and surface feedbacks which operate on 10 day time scale, but it should be more linked to the multiscale nature of atmospheric circulation and its faster effect in the stirring of humidity and vorticity fields. The recently observed presence of Shallow Mesoscale Overturning Circluations in the tropics \citep{george2023widespread}, is an evidence of the strong coupling between convection and multi-scale circulations, as also demonstrated by idealized LES simulations \citep{Janssens2024shallow}. In this latter study, mesoscale vertical motions were found to be in WTG (Weak Temperature Gradient) balance between 12.5 and 400 km scales and driven by convective heating, rather than by radiative cooling.

In this work, we analyze the occurrence and the driving mechanisms of multi-scale circulations in deep convective self-aggregation. To this aim, we introduce a scale-by-scale theoretical framework that is novel in the context of atmospheric sciences. Such framework is based on the generalized Yaglom equation \citep{yaglom1949local}, which describes the evolution of the scale-energy variance, that is the variance of energy associated to each scale. This formalism provides a full description of the strongly inhomogeneous (elevation dependent) and anisotropic (scale dependent) processes governing deep convection thus allowing to unveil mechanisms that would be otherwise missed by classical (elevation dependent) single-point statistics. This theoretical framework has already been adopted in previous studies to analyze scale-by-scale budgets in more simplified frameworks such as homogeneous turbulence \citep{gauding2014generalised}, grid turbulence \citep{danaila1999generalization}, Rayleigh-Benard convection \citep{rincon2006anisotropy, togni2015physical, togni2019resolved}, turbulent wakes \citep{antonia1997analogy} or in atmospheric boundary layer observations \citep{nowak2025scale}. We will apply this framework to the moist-static energy variable in line with the results obtained in part I of the present research activity based on single-point statistics \citep{silvestri2026gye_part1} and with the literature results based on the most common column-integrated variance framework \citep{wing2014physical}.

Although this work focuses on the scale-by-scale budget of moist static energy variance, a previous work by \citet{lai2023kinetic} studied the processes regulating the kinetic energy spectrum by applying a spectral budget framework on both aggregating and non-aggregating RCE simulations with the WRF model. They show that the kinetic energy spectrum of the aggregated simulation in the upper troposphere (6-10 km) resembles a -5/3 power law for wavelengths above 20 km, in agreement with spectra observed from aircraft data \citep{gage1985climatology,cho1999horizontal} and in contrast with steeper scaling (-11/5 power law) as expected from buoyancy-driven turbulence theory \citep{bolgiano1959spectra, obukhov1959effect}. However, since the nonlinear advective transfer term was relatively insignificant, the kinetic energy spectrum of aggregation was not interpreted as the result of a turbulent cascade, but rather as the result of a balance between the gain of energy by buoyancy flux (latent heating from convection) and the removal of energy by vertical energy fluxes. Such balance was fulfilled at wavelengths between 20 and 200 km. 
Another relevant study is that of \citet{yano2025emergence}  where the energy cycle of self-aggregation is analyzed using a wavelet-space decomposition of both kinetic energy and available potential energy in the Met Office Unified Model. Their results suggest that convective heating first generates available potential energy, which is then rapidly converted into kinetic energy through buoyancy forcing, consistent with the WTG approximation (to which they refer as free-ride principle). Indeed, their study found that this approximation, usually considered to characterize the large-scale balance of the tropical atmosphere \citep{sobel2001weak}, can be also extended to the convective scale. Then, a substantial fraction of this kinetic energy appears in the form of small-scale gravity waves propagating away from convective centers, and these waves can transfer energy upscale through nonlinear interactions, eventually sustaining a large-scale circulation associated with the organized precipitation pattern. The relevance of gravity-waves in the upscale growth of available potential energy structures shares similarities with previous work by \citet{yang2021shallow}. While \citet{lai2023kinetic} emphasizes that the aggregated state cannot be straightforwardly interpreted as the outcome of a turbulent kinetic-energy cascade, \citet{yano2025emergence} highlights an upscale transfer of wave-induced kinetic energy that supports the organized circulation.

Our scale-by-scale analysis based on the generalized Yaglom equation provides a complementary perspective to wavelet-based and spectral-based studies of tropical self-organization. Whereas a wavelet-space approach, such as that used by \cite{yano2025emergence}, is well suited to identifying the emergence, localization, and multiscale structure of coherent convective patterns, the generalized Yaglom framework offers an exact scale-by-scale budget that directly diagnoses the physical terms responsible for maintaining or weakening aggregation in RCE. In this sense, wavelet methods are particularly useful for describing where organized structures appear in scale space, while the Yaglom equation presented in this work is better suited to explaining why they persist or break down. Therefore, our goal is to quantify and isolate the dynamical mechanisms that sustain the stationary aggregated state and especially to describe the mechanisms that are generating, consuming and transporting moist static energy variance both in the physical space and in the scale space. The use of such a scale-by-scale framework will allow us to find relevant scaling and to derive reduced-order equations that are relevant for an improved description and modeling of self-aggregation in climate models.

The paper is organized as follows. In section \ref{sec:setup} the simulation setup is briefly described. In section \ref{sec:gye} the generalized Yaglom equation (GYE) is derived and specialized for the RCE idealized configuration. In section \ref{sec:single-point} we show the convergence of this framework at large-scales to the single-point variance budget already presented in part I of the present research activity \citep{silvestri2026gye_part1}. Then Sections \ref{sec:scale_var}, \ref{sec:source_scale_var} and \ref{sec:path_scale_var} show the main results about the distribution of scale-energy variance, the sources of scale-energy variance and the paths of scale-energy variance in the space of scales and physical positions, respectively. Section \ref{sec:phenomen} offers a phenomenological view of such energy paths and how they can be translated into the convective structures observed in physical space and reported in part I of the present research activity \citep{silvestri2026gye_part1}. The budget of scale-energy variance is shown for different atmospheric layers in Section \ref{sec:scale_budget} and a reduced-order model is derived for each layer. Finally, the scalings of second-order structure functions offer a self-similar view of mesoscale structures which are used to estimate the maximum size over which spontaneous aggregation processes can act. The conclusions are provided in Section \ref{sec:conclusions} and final discussions close the work in Section \ref{sec:disc}.

\section{Simulation setup}
\label{sec:setup}
The System of Atmospheric Modeling \citep[SAM version 6.10.5,][]{khairoutdinov2003cloud} is employed here to study convective self-aggregation. The main equations and assumption of such model, as well as the adpoted physical parametrizations, are reported in part I of this work \citep{silvestri2026gye_part1}. Here we briefly summarized the main features of the simulations.
The computational grid has uniform spacing in lateral direction and stretched spacing in the vertical with $N_x,N_y,N_z=(256,256,64)$. The resulting horizontal grid resolution is $\Delta x=\Delta y=3$ km, small enough to consider the simulation "convective and cloud permitting" with a domain size (768 km by 768 km) large enough to allow the spontaneous development of convection, based on previous RCE diagram on the same model \citep{muller2012detailed}. On the vertical, the grid spacing gradually increases from 50 m near the ground to 500 m above 5 km. Then it increases again from 500 m to 1 km above 20 km. The simulation is initialized by a uniform vertical profile of humidity and temperature, without no mean wind imposed. At the surface a uniform SST of about 302 K is kept constant throughout the simulation and the convection is initialized by a random perturbation of the temperature field over the lower model levels. We imposed lateral periodic boundary conditions, such that the calculated statistics are a function only of the vertical direction $z$. The average operator $\langle\, \cdot \,\rangle$ is used to denote an average both in time (last 10 days) and space (only horizontal directions). Instead, the average operator $\overline{\;\cdot\;}$ is used to denote only an average along the horizontal directions. The fluctuation of a certain quantity $Y(x,y,z,t)$ with respect to both space and time average is indicated as $Y^{\prime}(x,y,z,t)=Y(x,y,z,t)-\langle Y \rangle (z)$, while the fluctuation with respect only to space is indicated as $Y^{\prime\prime}(x,y,z,t)=Y(x,y,z,t)-\overline{Y}(z,t)$. Averages in time will be performed over the last 10 days, for a total of 240 samples.

\section{The generalized yaglom equation and the augmented space of deep convection}
\label{sec:gye}

In part I of the present research activity \citep{silvestri2026gye_part1}, we have seen that the flow in the self-aggregated configuration is organised into multi-scale structures, with one large-scale circulation involving a dry region and a moist region. The characteristic length of such large-scale circulation is about half domain length, whereas the steady deep convective cluster has a diameter of about 300 km. However, many small scale structures are evident especially in the boundary layer. Inside the boundary layer of the dry region, there are smaller scale dry and moist plumes. Eventually also shallow clouds are present at the top of such convective plumes. Similarly, inside the moist region, there are mesoscale structures like cold pools and deep convective towers, as well as congestus clouds (mid-level clouds). Around cold pools, convergence lines of deep convection can form with a typical scale of about 100 km. The large moist cluster of deep convection at scales of 300 km, is therefore associated to the merging of multiple convergence lines and cold pools which, on average, give rise to an inner region of lower $h_f$ anomaly and higher $h_f$ on its edge \citep[super cold pool,][]{windmiller2019convection}.

In order to characterize such multi-scale dynamics, we derive an equation for the frozen moist static energy variance that depends both on the physical space position $\vb{x_c}$, to take into account the strong flow inhomogeneity in the vertical direction, and the scale space $\vb{r}$, to take into account the different scales involved. The equation is a generalized form of the Yaglom Equation (GYE) \citep{yaglom1949local}. It represents the evolution equation for the second-order structure function of frozen moist static energy per unit volume,
\begin{equation}
    \rhobar\langle \delta h_f'^2 \rangle \equiv \rhobar \langle \delta h_f' \delta h_f' \rangle
    \label{eq:struct_func}
\end{equation}
where $\rho$ is the density of the fluid and the operators $\delta (\cdot)$ and $\widetilde{(\cdot)}$ denote the two-points increment and average that for a generic quantity $\gamma$ read
\begin{equation}
    \delta \gamma   \equiv \gamma (x_i^1,t)- \gamma (x_i^2,t) \qquad \widetilde{\gamma}   \equiv (\gamma (x_i^1,t) + \gamma (x_i^2,t))/2 \, .
\end{equation}
Finally, the operators $\langle \cdot \rangle$ and $\overline{(\cdot )}$ denote the ensemble and the horizontal averages and are technically described in part I of the present research activity \citep{silvestri2026gye_part1}. By introducing the separation vector $\vb{r}=\vb{x^1}-\vb{x^2}$, and the mid-point position vector, $\vb{x_c}=(\vb{x^1}+\vb{x^2})/2$, it is possible to highlight that the second-order structure function of frozen moist static energy (\ref{eq:struct_func}) depends on a seven-dimensional augmented space, $\langle \delta h_f'^2 \rangle = \langle \delta h_f'^2 \rangle (\vb{x_c},\vb{r},t)$. In particular, the two-point separation vector $\vb{r}$ defines the three-dimensional space of scales while the mid-point position vector $\vb{x_c}$ identifies the three-dimensional space of positions. Accordingly, $\rhobar\langle \delta h_f'^2 \rangle$ is a measure of the variations of the anomaly of frozen moist static energy per unit volume at different scales and positions and for that reason will be hereafter called scale energy variance.

The exact equation for the scale energy variance can be obtained from the evolution equations of the system by following the approach introduced in \citet{hill2002exact} for the velocity field in incompressible turbulent flows. In its most general form, the equation reads
\begin{align}
    \pdv{ \langle \delta h_f'^2 \rangle}{t} &+  \langle \widetilde{u_j'} \pdv{\delta h_f'^2}{x_{c_j}} \rangle +  \langle \delta u_j' \pdv{\delta h_f'^2}{r_j} \rangle +  \widetilde{U_j} \pdv{\langle \delta h_f'^2 \rangle}{x_{c_j}} + \delta U_j \pdv{\langle \delta h_f'^2 \rangle}{r_j} + \nonumber \\
    &+ 2 \langle \delta h_f' \widetilde{u_j'} \rangle \, \delta \left (\pdv{H}{x_j} \right ) + 2 \langle \delta h_f' \delta u_j' \rangle \widetilde{\left (\pdv{H}{x_j} \right )}  =\nonumber \\ 
    &=- L_f \left(\irhobarp + \irhobarpp \right) \langle \delta h_f' \pdv{\delta P_f'}{z_{c}} \rangle - 2 L_f \left (\irhobarp - \irhobarpp \right ) \langle \delta h_f' \pdv{\delta P_f'}{r_j} \rangle + \nonumber \\
    &- \left ( \irhobarp + \irhobarpp \right ) \langle \delta h_f' \pdv{\delta \tau_{h_f,j}'}{x_{c_j}} \rangle - 2 \left ( \irhobarp - \irhobarpp \right ) \langle \delta h_f' \pdv{\delta \tau_{h_f,j}'}{r_j} \rangle + 2 c_{pd} \langle \delta h_f' \delta Q_{rad}' \rangle
    \label{eq:gye_general}
\end{align}
where $H = \langle h_f \rangle$ and $U_i = \langle u_i \rangle$ denote the average fields of frozen moist static energy and velocity, $\tau_{h_f,j}$ are the subgrid scalar fluxes, $Q_{rad}$ is the radiative heating rate and $P_f$ are the frozen precipitation fluxes. For a more detailed definition of the transported quantities, the reader is referred to part I of the present research activity \citep{silvestri2026gye_part1}. The generalized Yaglom equation (\ref{eq:gye_general}) allows to study the scales and positions at which scale energy variance is produced and eventually dissipated. Furthermore, it allows to trace the paths followed by scale energy variance from its production to its dissipation in the augmented space of scales and positions thus enabling a complete multi-dimensional description of the evolution of the anomaly of frozen moist static energy. The complete derivation of the generalized Yaglom equation (\ref{eq:gye_general}) is reported in the appendix.

By considering the statistical symmetries of the radiative-convective equilibrium problem, the dimensionality of the problem described by the generalized Yaglom equation drastically reduces. In particular, because of statistical steadiness of the final deep convective self-aggregation state, the dependence on time is dropped. Because of statistical homogeneity in the horizontal directions the space of positions $\vb{x_c}$ reduces to the one-dimensional space of atmospheric levels $z_c$. Finally, because of statistical isotropy in the horizontal directions, the space of scales $\vb{r}$ reduces to the two dimensional space $\left (\sqrt{r_x^2+r_y^2}, r_z \right)$. Hence, scale energy variance for the radiative-convective equilibrium problem is defined in a three-dimensional space of scales and elevations,
\begin{equation}
\langle \delta h_f'^2 \rangle = \langle \delta h_f'^2 \rangle \left (r, r_z, z_c \right )
\label{eq:strc_func_rce}
\end{equation}
with $r= \sqrt{r_x^2+r_y^2}$. Within this statistical framework and by reducing the analysis of the space of scales to the sole space of horizontal scales $r$ for $r_z = 0$, the generalized Yaglom equation (\ref{eq:gye_general}) multiplied by $\rhobar(z_c)$ drastically simplifies and reduces to the following form,
\begin{align}
    \pdv{\rhobar \langle \delta h_f'^2 \widetilde{w'} \rangle}{z_c} &+ \pdv{\rhobar \langle \delta h_f'^2 \delta u_j' \rangle}{r_j} \rangle + 2 \rhobar \langle \delta h_f' \delta w' \rangle \left(\pdv{H}{z} \right)  \nonumber \\
    &= - 2 L_f \langle \delta h_f' \, \delta \left (\pdv{P_f'}{z} \right ) \rangle  -4\pdv{\langle \delta h_f' \widetilde{\tau_{h_f,j}'} \rangle}{r_j}  - \pdv{\langle \delta h_f' \delta \tau_{h_f,z}' \rangle}{z_c} - 4 \langle \epsilon_{sgs}^h \rangle + 2 \rhobar c_{pd} \langle \delta h_f' \delta Q_{rad}' \rangle
    \label{eq:gye_RCE}
\end{align}
where $\epsilon_{sgs}^h = - \tau_{h_f,j}' (\partial h_f'/\partial x_j)$ is the flux of frozen moist static energy from/to subgrid scales. In symbolic form, equation (\ref{eq:gye_RCE}) can be rewritten as
\begin{equation}
    T_r+D_r+T_c+D_c+\Pi +Q-E +R = 0
    \label{eq:gye_symb}
\end{equation}
where it is possible to highlight the three main physical processes described by the generalized Yaglom equation. A scale energy variance transport among different scales $(T_r + D_r)$ performed by inertial mechanisms such as circulations and advection,
\begin{equation}
    T_r= -\pdv{\rhobar  \langle \delta h_f'^2\delta u_j' \rangle}{r_j}
\end{equation}
and by diffusion,
\begin{equation}
    D_r=-4\pdv{\langle \delta h_f' \widetilde{\tau_{h_f,j}'} \rangle}{r_j} \, .
\end{equation}
A scale energy variance transport among different atmospheric levels $(T_c + D_c)$ performed by inertial mechanisms
\begin{equation}
    T_c= - \pdv{\rhobar  \langle \delta h_f'^2 \widetilde{w'} \rangle}{z_c} 
\end{equation}
and by diffusive processes
\begin{equation}
    D_c= - \pdv{\langle \delta h_f' \delta \tau_{h_f,z}' \rangle}{z_c} \, .
\end{equation}
Finally, scale energy variance source and sink mechanisms $(\Pi +Q -E + R)$ that are local in the space of scales and positions and that are given by the statistical inhomogeneity (gradient) of the mean frozen static energy,
\begin{equation}
    \Pi = -2 \rhobar  \langle \delta h_f' \delta w' \rangle \left(\dv{H}{z} \right)
\end{equation}
by radiation,
\begin{equation}
    Q=2 \rhobar c_{pd} \langle \delta h_f' \delta Q_{rad}' \rangle 
\end{equation}
by subgrid dissipation,
\begin{equation}
    E= 4 \langle \epsilon_{sgs}^h \rangle
\end{equation}
and by the frozen precipitation fluxes,
\begin{equation}
    R=- 2 L_f \langle \delta h_f' \, \delta \left (\pdv{P_f'}{z} \right ) \rangle \, . 
\end{equation}

In order to highlight the field of fluxes in the compound space of scales and atmospheric levels, it is useful to rewrite equation (\ref{eq:gye_RCE}) in a conservative form
\begin{equation}
    \grad \boldsymbol{\cdot} \vb*{\phi}=\zeta
    \label{eq:gye_flux}
\end{equation}
where the source/sink terms of scale energy variance have been grouped together in a total source term,
\begin{equation}
    \zeta=\Pi +Q-E +R \, .
\end{equation}
and, in order to highlight the statistical isotropy of the space of horizontal scales, equation (\ref{eq:strc_func_rce}), a cylindrical reference frame is considered for the space of scales so that 
\begin{equation}
    \grad =[\grad_r, \grad_c]=\left [\frac{1}{r}\pdv{r()}{r},\pdv{()}{r_z},\pdv{()}{z_c} \right]
\end{equation}
where $r = \sqrt{r_x^2+r_y^2}$. The conservative form of the generalized Yaglom equation (\ref{eq:gye_flux}) highlights that a three dimensional flux of scale energy variance occurs,
\begin{equation}
    \vb*{\phi}=    
    \begin{bmatrix}
    \phi_{r} \\
    \phi_{r_z} \\
    \phi_{c_z} 
    \end{bmatrix} =
    \begin{bmatrix}
    \rhobar  \langle \delta h_f'^2\delta u_r' \rangle + 4 \langle \widetilde{\tau_{h_f,r}'} \delta h_f' \rangle \\
    \rhobar  \langle \delta h_f'^2\delta w' \rangle + 4 \langle \widetilde{\tau_{h_f,z}'} \delta h_f' \rangle \\
    \rhobar  \langle \delta h_f'^2 \widetilde{w'} \rangle+ \langle \delta \tau_{h_f,z}' \delta h_f' \rangle 
    \end{bmatrix}
\end{equation}
that is composed by a two-dimensional flux in the space of scales $(\phi_{r}, \phi_{r_z})$ and by a one-dimensional flux $\phi_{c_z}$ among different atmospheric levels. Here, $\delta u_r$ and $\tau_{h_f, r}$ denote the radial components of the velocity and subgrid diffusion fields, respectively. Since, equation (\ref{eq:gye_flux}) together with equation (\ref{eq:gye_RCE}) are written for an hyperplane $r_z= 0$, only the two-dimensional field of fluxes $(\phi_r, \phi_{c_z})$ can actually be traced and analysed. In this framework, the divergence of the fluxes among vertical scales $\partial\phi_{r_z}/\partial r_z|_{r_z=0}$ becomes a net source/sink of scale energy variance related to the energy exchange with the space of $r_z \neq 0$ scales. Accordingly, an extended source term can be defined
\begin{equation}
    \xi =\zeta -\pdv{\phi_{r_z}}{r_z}
    \label{eq:xi_rz0}
\end{equation}
and the conservative form of the generalized Yaglom equation describing the evolution of the two-dimensional field of fluxes among horizontal scales and atmospheric levels $(\phi_r, \phi_{c_z})$ finally reads,
\begin{equation}
    \frac{1}{r}\pdv{(r \phi_r)}{r} +\pdv{\phi_{c_z}}{z_c} = \xi
    \label{eq:phi_rz0}
\end{equation}
Equation (\ref{eq:phi_rz0}) defines a theoretical framework that unequivocally identifies and quantifies phenomena of forward or reverse cascade of energy that could be at the basis of the deep convective self-aggregation state where a large organized structure emerges from small scale structures. In particular, forward/reverse cascade denotes fluxes in the space of horizontal scales $r$ that move energy towards smaller and larger scales, corresponding to $\phi_{r}<0$ and $\phi_{r}>0$, respectively. On the other hand, descending/ascending spatial fluxes denote subsidence/convective fluxes in the physical space that move energy towards lower and higher atmospheric levels $z_c$, corresponding to $\phi_{c_z}<0$ and $\phi_{c_z}>0$, respectively.

\section{Asymptotic behaviour and single-point energy variance budget}
\label{sec:single-point}

\begin{figure}[t]
    \centering
    \includegraphics[width=0.6\linewidth]{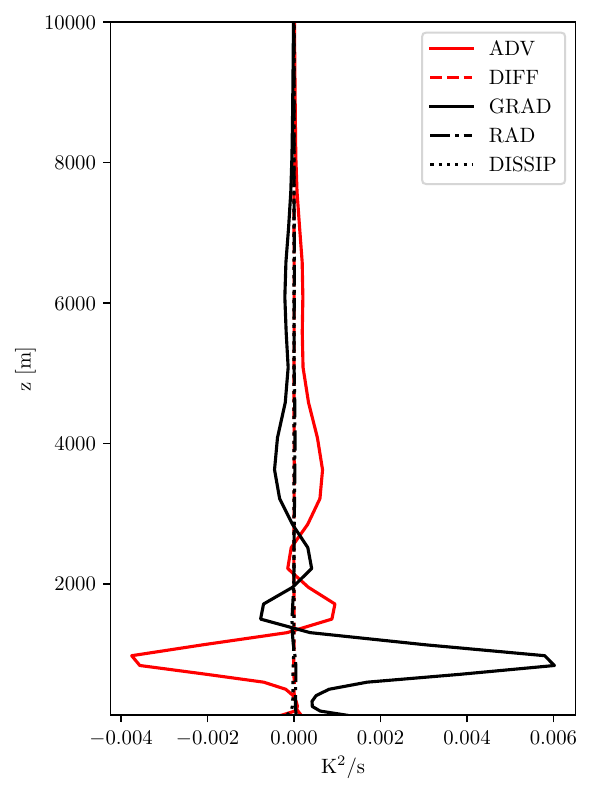}
    \caption{Large-scale asymptotic behaviour of the generalized Yaglom equation (\ref{eq:gye_RCE}) evaluated for very large scales, $r \approx 380Km$. The recovery of the single-point budget (\ref{eq:var}) can be appreciated by comparing with results from part I of the present research \citep{silvestri2026gye_part1}.}
    \label{fig:var_single_point}
\end{figure}

At scales larger than the correlation length, the points $\vb{x^1}$ and $\vb{x^2}$ become uncorrelated \citep{marati2004energy} and the generalized Yaglom equation (\ref{eq:gye_RCE}) reduces  within a factor of 4, to the single-point energy variance budget presented in part I of the present research \citep{silvestri2026gye_part1}: 
\begin{align}
    \underbrace{\dv{\rhobar \langle w' h_f'^2\rangle}{z}}_{ADV}&+ \underbrace{2\dv{\langle\tau_{h_f,z}'h_f'\rangle}{z}}_{DIFF}= \nonumber \\
    &=\underbrace{-2\overline{\rho}\langle w' h_f'\rangle\dv{H}{z}}_{GRAD} \underbrace{-2L_f\langle h_f'\dv{P_f'}{z}\rangle}_{PREC}-\underbrace{2\langle\epsilon^{h}_{sgs}\rangle}_{DISSIP} + \underbrace{2\rhobar c_{pd} \langle h_f' Q_{rad}'\rangle}_{RAD}
    \label{eq:var}
\end{align}
As shown in figure \ref{fig:var_single_point}, the generalized Yaglom equation (\ref{eq:gye_RCE}) evaluated for very large scales ($r \approx 380$ km) asymptotically recovers the results of the single-point budget (\ref{eq:var}) reported in figure 12(b) of part I of the present research \cite{silvestri2026gye_part1}.

The behaviour of the single-point budget is relevant for a better understanding the complex multi-dimensional results provided by the generalized Yaglom equation. Indeed, it allows to characterize the vertical inhomogeneity of the flow by identifying the relevant regions composing the atmospheric layer. For this reason, we briefly summarize here the main results of the single-point budget equation (\ref{eq:var}) presented in part I of the present research \citep{silvestri2026gye_part1}. The analysis reveals a layered structure where source/sink of variance by the gradient term is mainly balanced by the vertical transport of variance. The different layers identified are:
\begin{enumerate}
    \item the near-surface layer, between 0 and 250 m, where subgrid fluxes play a significant role and a first positive peak of production by the gradient term occurs;
    \item the boundary layer, between 250 m and 750 m, a well mixed region where $\langle h_f \rangle$ is almost homogeneously distributed thus leading to small values of $dH/dz$ and, hence, to a strong decrease of production;
    \item the boundary layer top, between 750 m and 1200 m, where we find the strongest gradient of $\langle h_f \rangle$ and shallow clouds start to form and a strong radiative cooling is present which drives the largest production of variance;
    \item the shallow clouds layer, between 1200 m and 2000 m, where we find a sink of variance due to the gradient term and the increase of $\langle h_f \rangle$ with height;
    \item the lower troposphere above the shallow clouds, between 2000 m and 3000 m, where $\langle h_f \rangle$ decreases again with height and a secondary peak of production is found; 
    \item the mid free troposphere, from 3500 to 8000 m, a sink region where the gradient of $\langle h_f\rangle$ is negative and almost constant;
    \item the upper troposphere above 8000 m, where substantially no variance contributions are found;
\end{enumerate}

\section{Scaling of energy and velocity variance in the augmented space $(r, z_c)$}
\label{sec:scale_var}

\begin{figure}[t]
    \centering
    \includegraphics[width=0.8\linewidth]{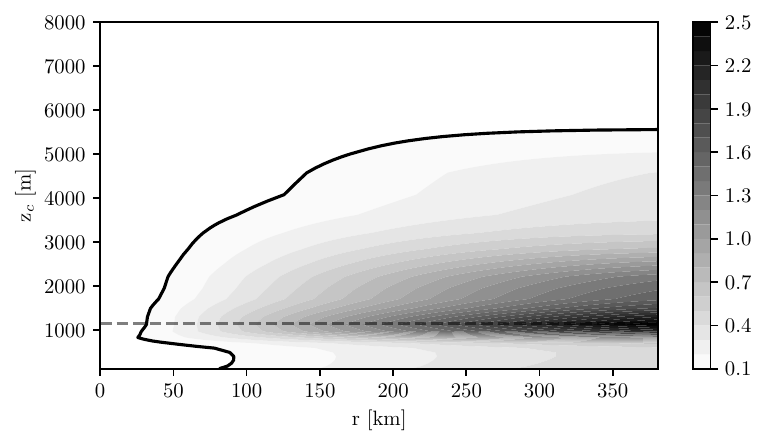}
    \caption{Contour of scale-energy variance $\langle \delta h_f'^2 \rangle/\langle h_f'^2\rangle_{max}$ in the multi-dimensional space $(r,z_c)$. The horizontal dashed line shows the boundary layer top at 1150 m above the surface where the maximum energy variance is found. The thick black line reports the isoline of $\langle \delta h_f'^2 \rangle/\langle (h_f'^2) \rangle_{max} = 0.1$. Note that aspect ratio between horizontal and vertical scale is not unity, but the vertical dimension is elongated for visibility.}
    \label{fig:scale_var}
\end{figure}

The scale-energy variance, $\langle \delta h_f'^2 \rangle$ is reported in Figure \ref{fig:scale_var} normalized by the peak of variance at the boundary layer top as found in Figure \ref{fig:var_single_point}. Most of the scale variance is concentrated at large scales and it is linked to the large-scale circulation that is present between dry and moist regions. The peak of variance is located in the boundary layer top region around 1150 m, coincident with the largest peak of the single-point scale variance. Going to smaller scales, the variance is more related to (turbulent) convective structures, such as deep convective cores and cold pools with a largest peak always in the boundary layer top.

In order to highlight the scales involved, the isoline $\langle \delta h_f'^2 \rangle/\langle (h_f'^2) \rangle_{max} = 0.1$ is also reported in Figure \ref{fig:scale_var}, whose position in the space of horizontal scales $r$ can be regarded as the turbulent convective length of scale-energy variance at different atmospheric levels $z_c$. The convective scale is almost constant in the near surface layer and boundary layer region and it is about 100 km. Therefore in the boundary layer region below 750 meters, the scales carrying most of the energy variance are of the order of 100 km. Such structures can be linked to the cold pools, where large regions of divergence with lower anomalies of $h_f$ are present at the surface and organize the convective cores with larger humidity in large arcs of length of the order of 100 km. As demonstrated by \cite{muller2012detailed} and \cite{Yanase2020critical} with two different models, simulations of convection with domain size of the order of 100 km, will not exhibit self-aggregation for any model and horizontal grid resolution. In such small domains, we cannot correctly represent the structures carrying most of the energy variance inside the boundary layer.

By moving to the boundary layer top, the turbulent convective length highlighted by the isoline $\langle \delta h_f'^2 \rangle/\langle (h_f'^2) \rangle_{max} = 0.1$ is found to decrease, reaching its minimum value of 27 km at about 850 m height. This means that the relevant structures for the energy variance have decreased their horizontal scale in this region with respect to the near-surface layer (stretching of large surface structures). From this minimum, the convective structures are found to significantly increase their length scale by increasing the elevation from the boundary layer top to the shallow cloud layer and to the lower and mid troposphere at about 6000 m where their lengths saturate to the numerical domain size. Accordingly, no relevant length scales can be identified above 6000 m thus highlighting the almost absence of mixing phenomena due to turbulence in the upper troposphere. This is also confirmed by the negligible intensity of energy variance above this atmospheric level.

\begin{figure}[t]
    \centering
    \includegraphics[width=1\linewidth]{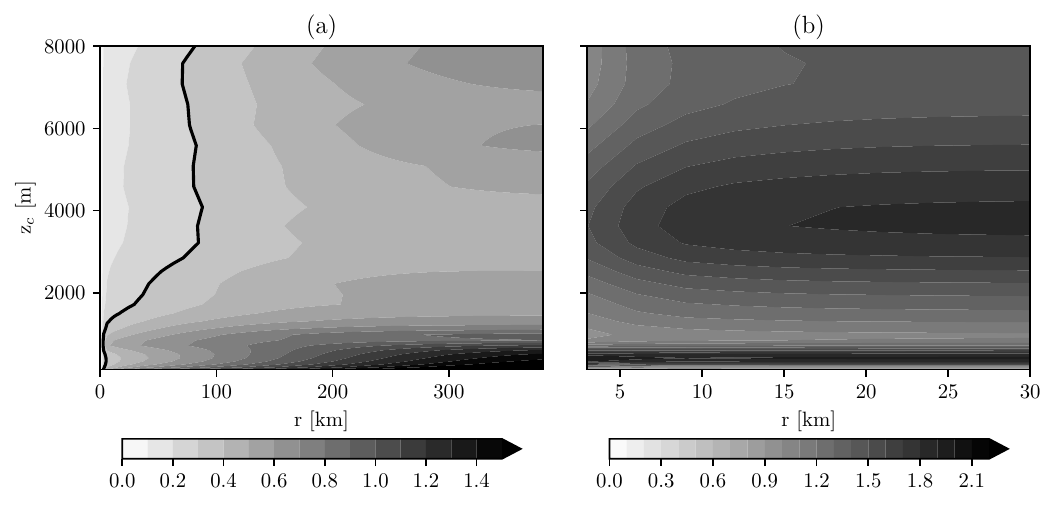}
    \caption{Contour of the second-order structure function of a) the horizontal velocity $\langle \delta u_r'^2 \rangle/\langle u_r'^2\rangle_{max}$ and b) the vertical velocity $\langle \delta w'^2 \rangle/\langle w'^2\rangle_{max}$ in the multi-dimensional space $(r,z_c)$. The tick black line in panel a, reports the value $\langle \delta u_r^{\prime 2} \rangle / \langle u_r^{\prime 2} \rangle_{max} = 0.3$. Note that the aspect ratio between horizontal and vertical scale is not unity, but the vertical dimension is elongated for visibility. In Figure (b) only the structure function at small scales is reported since for scales larger than 30 km the scale-energy variance is constant.}
    \label{fig:scale_var_du2}
\end{figure}

To better understand the pattern shown by the scale-energy variance, we consider now the behaviour of the velocity field by addressing the second-order structure function of the horizontal and vertical velocity components, $\langle \delta u_r^{\prime 2} \rangle$ and $\langle \delta w'^2 \rangle$ respectively. As shown in Figure \ref{fig:scale_var_du2}a, the regions of most intense activity of horizontal velocity fluctuations are the surface and boundary layers. A second weaker region of activity is found to be the mid troposphere above 4500 m. Note that the vertical locations of these two regions of activity of horizontal velocity fluctuations correspond to the inflow and outflow of deep convection. However, the scales involved are very different, thus highlighting a significant dissimilarity on the phenomena involved. As shown by the isoline $\langle \delta u_r^{\prime 2} \rangle / \langle u_r^{\prime 2} \rangle_{max} = 0.3$ reported in Figure \ref{fig:scale_var_du2}a, the horizontal velocity fluctuations in the surface and boundary layers are found to be active at scales that are significantly smaller than those attained in the troposphere. In particular, we measure that horizontal turbulence mixing is effective down to scales that on average are of the order of the grid scale size $r = 3$ km within the surface and boundary layers with the smallest values reached at the surface and at the boundary layer top. Such high values in the boundary layer are associated to existing shallow local scale circulations in both moist and dry regions linked to downdraft and updraft dynamics, as will be shown in Section \ref{sec:phenomen}. These scales increase from the top of the boundary layer saturating to values of the order of $r = 80$ km in the mid troposphere for $z_c \ge 4000$ m. Overall, this scenario further confirms the efficiency of mixing phenomena in the surface and boundary layers with respect to those in the troposphere. Furthermore, the sharp increase of the scales of horizontal mixing just above 850 m together with the already observed change in the convective scales for scale-energy variance at the same atmospheric levels, suggests that the boundary layer top can be interpreted as a sliding and deformable atmospheric layer that separates very different dynamics governing the well mixed lower atmospheric levels from the larger convective structures of the troposphere.

The activity of vertical velocity fluctuations in the compound space of scales and surface distances $\langle \delta w'^2 \rangle$ is shown in Figure \ref{fig:scale_var_du2}b. Compared to the horizontal velocity fluctuations, a completely different behaviour is observed, thus highlighting a strong anisotropy in the dynamic of the atmosphere. In particular, the scale-space behaviour of $\langle \delta w'^2 \rangle$ is found to saturate at scales significantly smaller than those of $\langle \delta u_r'^2 \rangle$, around $r \sim 30$ km. This simple feature suggests that the mixing action of the velocity field is performed by flow structures where vertical convection is coherent over shorter lateral distances with respect to lateral convection. This is also the reason why Figure \ref{fig:scale_var_du2}b reports statistics only for length scales smaller than 30 km. This saturation length is in agreement with similar statistics computed by \cite{islam1993predictability} and \cite{bongioannini2005orographic} on the precipitation field, where it is shown that convection is in quasi-equilibrium with scales larger than 30 km.

A more detailed analysis of Figure \ref{fig:scale_var_du2}b reveals that vertical convection is mainly active in two regions. The first is located right above the surface layer and occupies the boundary layer region, with an intensity which is constant for all scales, thus highlighting that the most intense convective structures populating these low atmospheric levels occur at scales of the order of the numerical resolution. The second peak activity of vertical convection is located at around 4000 m in the mid-troposphere. This is the location where the horizontal velocity fluctuations show a minimum value thus supporting the idea of the presence of large-scale circulations involving the entire low and mid troposphere. By considering that vertical convection is active at scales that are significantly smaller than those where lateral convection is active in the boundary layer top and in the mid troposphere (compare Figures \ref{fig:scale_var_du2}a and b), it is also possible to argue that the large-scale circulations of the troposphere are actually composed by a cluster of small scale structures of vertical convection.

In closing this section, let us anticipate that the different scalings of energy and velocity variances in the compound space of scales and elevations, $\langle \delta h_f'^2 \rangle$, $\langle \delta u_r'^2 \rangle$ and $\langle \delta w'^2 \rangle$, will be used in Section \ref{sec:self-similar-scaling} to derive self-similar scaling and theoretical estimates for RCE.

\section{Sources and sinks of scale energy variance}
\label{sec:source_scale_var}

The behaviour of scale energy variance shown in the previous section is the result of dynamical processes that generate, transport and finally dissipate anomalies of frozen moist static energy in the flow system. The generalized Yaglom equation (\ref{eq:gye_RCE}), especially its conservative form (\ref{eq:gye_flux}), describes in an unequivocal way all these processes in the augmented space of scales and elevations. In this section, we start the analysis by addressing the physical processes that generate/dissipate the anomalies of frozen moist static energy.

\begin{figure}[t]
    \centering
    \includegraphics[width=0.9\linewidth]{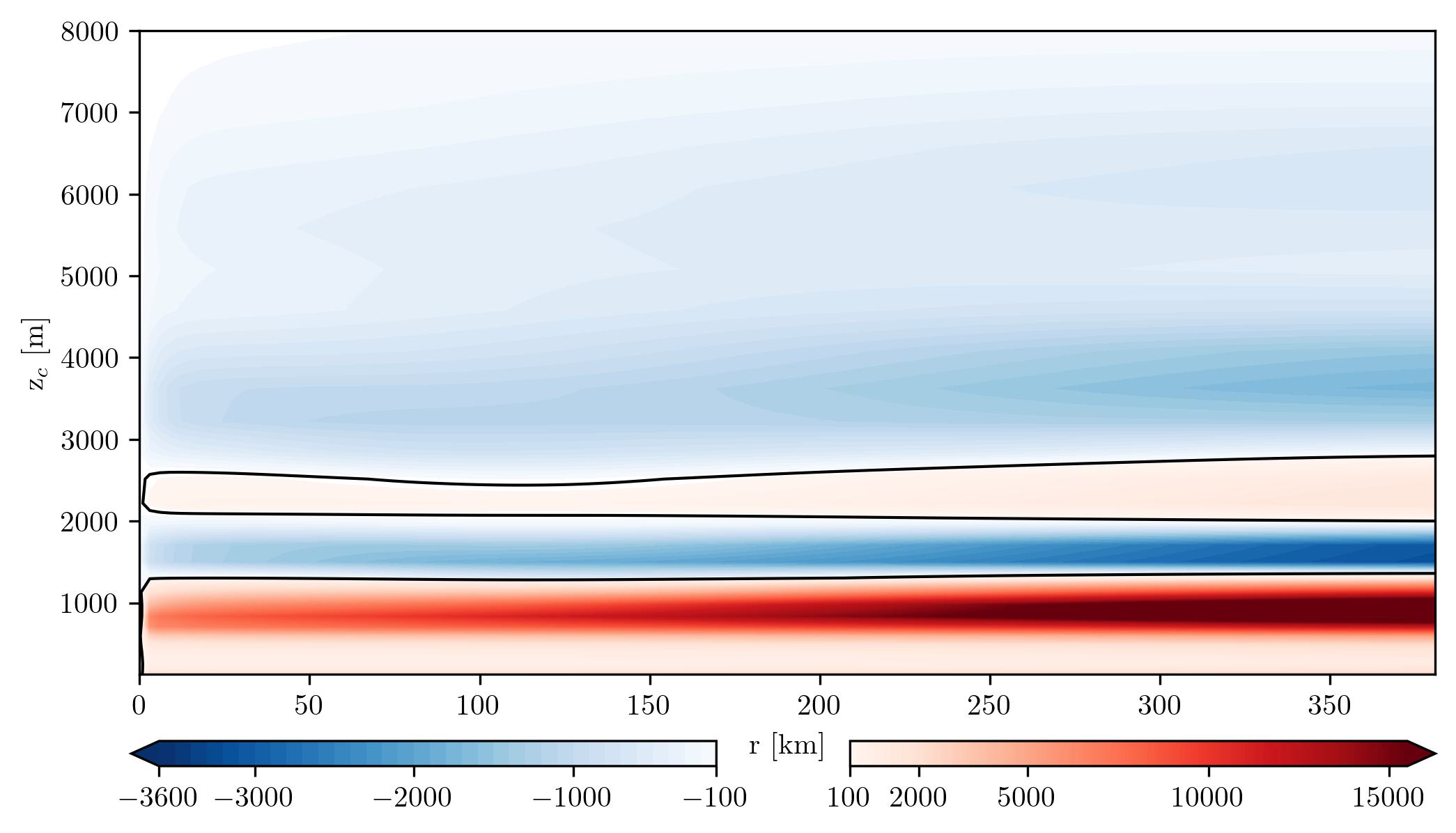}
    \caption{Isocontours of the extended source term $\xi$ as a function of horizontal scale $r$ and height $z_c$. Red colors indicate regions where $\xi>0$ (source of energy), while blue colors indicate regions where $\xi <0$ (sink of energy). The solid black line reports the isoline $\xi=0$.}
    \label{fig:source}
\end{figure}

The behaviour of the extended source/sink term $\xi$ in the augmented space of horizontal scales $r$ and heights $z_c$, is shown in Figure \ref{fig:source} with isocontours. Red colors indicate that $\xi>0$ acting as a source of energy, while blue colors indicate that $\xi <0$ acting as a sink of energy. The cross-over between source and sink is shown with a solid black line corresponding to the isoline $\xi=0$. Starting from the surface up to the boundary layer top region (1200 m), a large energy source is present, with a local maximum occurring at about 900 m, just above the end of the well-mixed boundary layer where the energy variance production by the gradient term reaches a maximum (Figure \ref{fig:var_single_point}), and just below the local maxima of the scale energy variance at 1150 m. This production peak is a large scale phenomenon and decreases its magnitude by moving towards small scales. On the other hand, the source of scale energy variance in the near surface layer is found to be almost equally distributed among scales, thus highlighting that also small scale structures participate in the production of FMSE anomalies in this region. Above 1200 m up to 2000 m, the shallow clouds layer is found to be a sink layer of scale energy. The maximum of this energy sink is found at about 1600 m. Interestingly, the energy sink in the shallow clouds layer increases in magnitude with scales. Indeed, this sink is given by the change of sign of the gradient term, $\Pi<0$ that is a large-scale phenomenon in nature contrary to diffusive term $E$ that on the contrary is a small-scale phenomenon. Above such a sink layer, in the lower free troposphere, another source region is present from 2000 m up to 3000 m with a smaller magnitude than the boundary layer source region. The atmospheric levels occupied by this secondary source region are actually thinner at the mesoscale with respect to the local and largest scales. This feature suggests that two separate large and small scale mechanisms participate to the production of FMSE anomaly at the lower free troposphere levels. From 3000 m to 8000 m, where the scale energy variance becomes very small, a weak sink region extends at every scale with two relative peaks of magnitude located at 3200 m and 6000 m. Again, large-scales are the most dissipative ones, being the sink driven by the inertial mechanisms represented by the gradient term $\Pi<0$.

Overall, the sources and sinks of scale energy alternate each other with a layered structure and generally increase their magnitude toward larger scales. Except for the surface layer up to 250 m above the surface, where radiation, dissipation and sub-grid scale diffusion are relevant for scale-energy production/dissipation, at larger distance from the surface the source/sink of scale energy variance results mainly from the contribution of the gradient term and the transfer term along the $r_z$ scale space. Hence, a reduced form for the source/sink term of the generalized Yaglom equation can be written as:
\begin{equation}
    \xi \approx \Pi + T_{r_z} = -2 \rhobar  \langle \delta w' \delta h_f' \rangle \left (\dv{H}{z} \right) - \pdv{ \langle \delta h_f'^2\delta w' \rangle}{r_z}
\end{equation}
with the gradient term being the dominant term in the primary source region. The increase of production with scale is related to the vertical flux of scale-energy increment $ \langle \delta w' \delta h_f' \rangle$, which is larger at the scales of atmospheric circulation (larger horizontal increment of both $\delta h_f'$ and $\delta w'$), than at the smaller scale structures.

\section{Paths of scale-energy variance in the compound space of scales and elevations}
\label{sec:path_scale_var}

In accordance with the generalized Yaglom equation in its conservative form (\ref{eq:gye_flux}), a field of fluxes driven by inertial and diffusive mechanisms establishes in the space of scales and of atmospheric levels. This field connects the source regions of the augmented space where an excess of scale energy variance is produced with the sink regions where it is eventually dissipated. Overall, the field of fluxes and the location of the source and sink regions allow us to have a complete view of the physical processes governing deep convection, in particular those phenomena establishing the field of scale energy variance analysed in the previous Section \ref{sec:scale_var}.

\begin{figure}[htbp]
    \centering
    \includegraphics[width=0.9\linewidth]{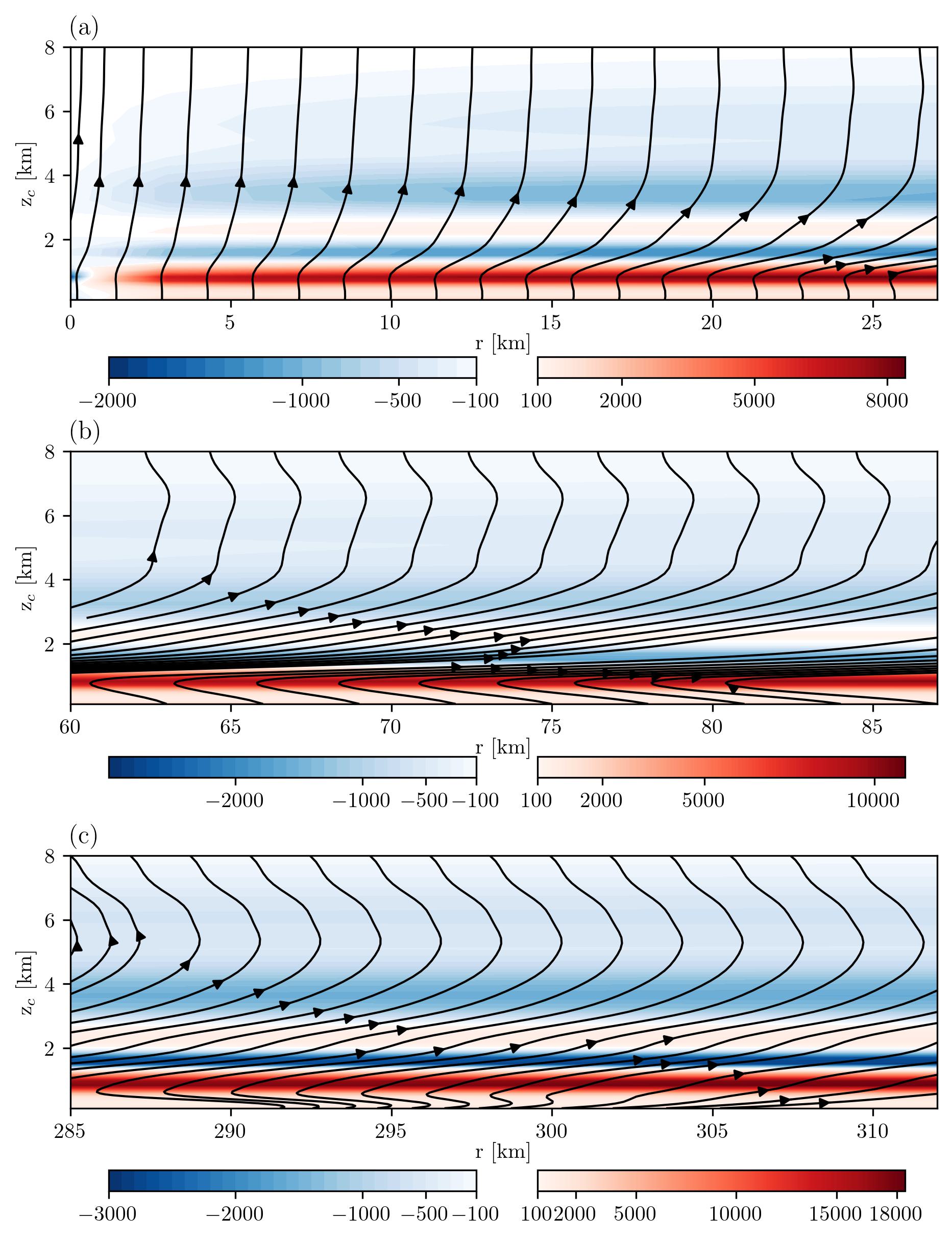}
    \caption{Fluxes of scale energy (black lines and arrows) in the compound space of scales and distance from the surface $(\phi_{r},\phi_{c_z})(r,z_c)$ with $r_z=0$ and isocontours of the source of scale energy $\xi(r,z_c)$ for: (a) local scales from 0 to 30 km, (b) mesoscale structures like cold pools and deep convective aggregates from 60 km to 90 km, and (c) cluster scale from 285 km to 315 km.}
    \label{fig:path}
\end{figure}

The field of fluxes of scale energy in scale and physical space $(\phi_r, \phi_{c_z})$ is reported in Figure \ref{fig:path} with streamlines. The fluxes are super-imposed to the iso-contour of the extended source term, $\xi$. An equal aspect ratio between vertical and horizontal scales is used in order to not distort the streamlines of the field of fluxes. For this reason, we are forced to split the result in three figures corresponding to three representative range of scales: local scales from 0 to 30 km (Figure \ref{fig:path}a), mesoscales from 60 km to 90 km (Figure \ref{fig:path}b) and cluster scales from 285 km to 315 km (Figure \ref{fig:path}c). To give a more quantitative view, Figure \ref{fig:path_average} shows the behaviour in the space of scales of various terms of the generalized Yaglom equation evaluated at fixed heights.

\subsection{Spatial fluxes}

\begin{figure}[t]
    \centering
    \includegraphics[width=1\linewidth]{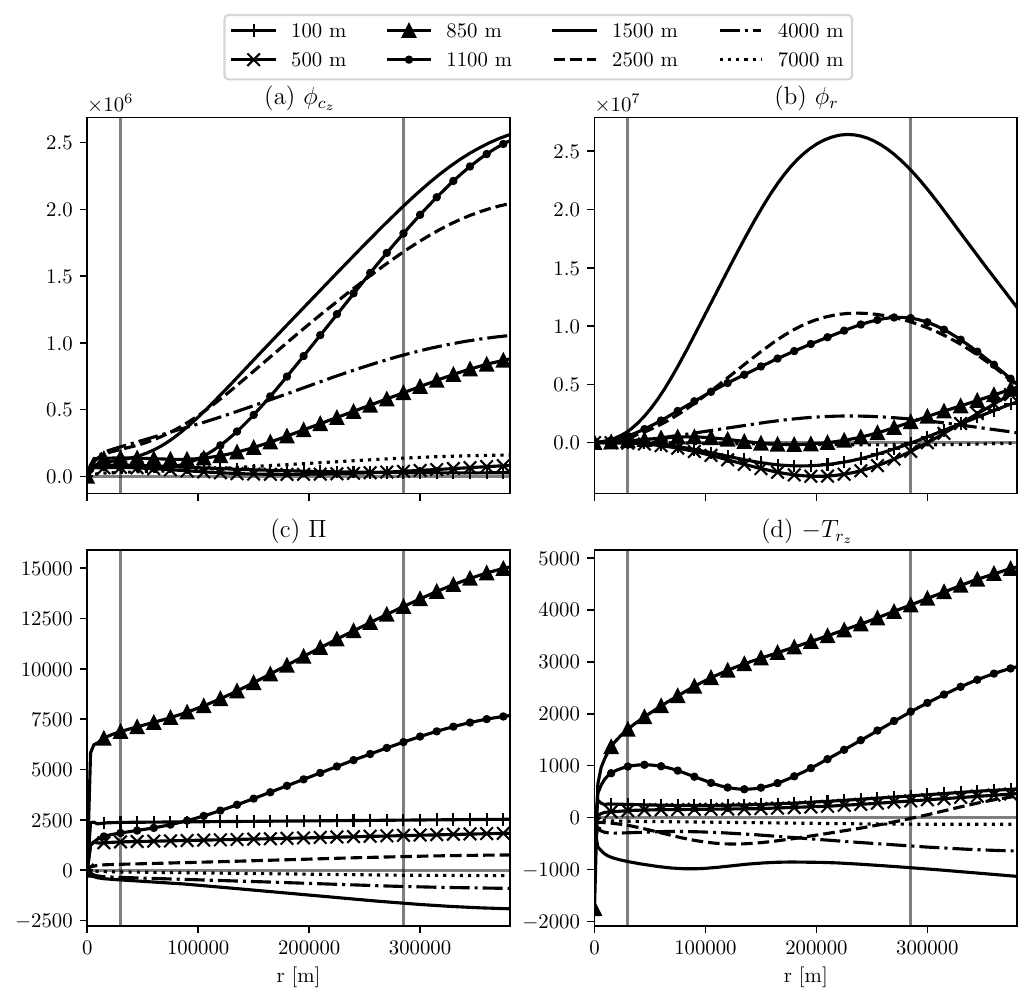}
    \caption{Horizontal scale dependence of energy fluxes and sources/sink at different distances from the surface: a) $\phi_{c_z}$ flux in the physical space; b) $\phi_r$ flux in the scale-space; c) $\Pi$ gradient source/sink term; d) additional source/sink term due to $T_{r_z}$ scale-space transport in the $r_z$ direction.}
    \label{fig:path_average}
\end{figure}

As shown in Figure \ref{fig:path}, the fluxes of scale-energy variance are always ascending, $\phi_{c_z}>0$, thus highlighting that a net vertical flux of scale-energy variance is traversing all scales of the entire atmosphere.  This flux is responsible for transporting anomalies of frozen moist static energy from the surface to the upper troposphere (see also Figure \ref{fig:path_average}a). As better quantified in Figure \ref{fig:path_average}a, this ascending spatial flux $\phi_{c_z}$ shows a modulation in the space of scales. In particular, it generally increases by moving towards larger scales, especially above the boundary layer region. The scale above which the spatial flux is increasing depends on the distance from the surface: in particular, in the boundary layer top this cross-over scale is of the order of 100 km while in the free troposphere is lower about 30 km. This feature highlights that almost all the structures of the free troposphere are significantly influenced by convection of FMSE anomaly, whereas only a portion of the phenomena populating the surface and boundary layers, the largest ones, are impacted by convection.

Starting from the ground and moving in the vertical direction, the spatial flux is found to increase in the surface and boundary layers at all scales (compare the different curves reported in \ref{fig:path_average}a). This increase is driven by the source nature of scale-energy variance of these two layers, $\xi > 0$. Indeed, the surface and boundary layers are the site of the largest production phenomena of FMSE anomaly $\xi_{max}$ whose action is to feed the vertical transport. Consequently, the spatial flux $\phi_{c_z}$ increases in these two regions and reaches a maximum by entering the sink region of the shallow cloud layer (1500 m). There, the spatial flux is not anymore sustained by local production mechanisms of FMSE anomaly, $\xi < 0$, rather it releases scale-energy variance thus reducing its intensity. Hence, the FSME anomalies of the shallow cloud layer are sustained by those formed below at the boundary layer top rather than being locally generated. By continuing its travel in the vertical direction, the spatial flux enters the lower troposphere between 2000 m and 3000 m. Although weaker, this region is again a source of FMSE anomaly $\xi > 0$. Hence, scale-energy variance there is sustained by local production mechanisms rather than by the spatial flux. The latter is indeed fed by the sourcing mechanisms of the lower troposphere. By moving upward, the mid and upper troposphere turn out to be sink regions of scale energy variance. Consequently, the spatial flux is no longer sustained by local production mechanisms and decreases its intensity (compare the different curves reported in \ref{fig:path_average}a). Hence, the FMSE anomaly in the mid and upper troposphere is provided by convective mechanisms transporting an excess of scale-energy variance provided by the production mechanisms of the lower troposphere and partially also of the surface and boundary layers.

\subsection{Scale-space fluxes in the near-surface and boundary layers}

Compared to the spatial flux $\phi_{c_z}$, a more complex behaviour is instead observed for the flux in the space of scales $\phi_{r}$. We start the analysis by considering the lower levels of the atmosphere, the surface and boundary layers between 0 m and 750 m. We recall that this region is a source region that is dominated by local production mechanisms of scale-energy variance that feed the ascending spatial flux towards the higher atmospheric levels. As shown in Figure \ref{fig:path}a, the scale space flux is very weak in the small scale range (local scales between 0 km and 30 km). As a result, the compound flux $(\phi_{r},\phi_{c_z})(r,z_c)$ is predominantly vertical, $|\phi_{r}|/|\phi_{c_z}| \approx 0$. By moving at mesoscales (Figure \ref{fig:path}b), the scale-space flux activates and a forward cascade of scale-energy variance $\phi_{r} < 0$ is observed, reaching its maximum strength at about 500 m elevation for mesoscales about 200 km, as better shown in Figure \ref{fig:path_average}b. In this region, the combination of a forward cascade $\phi_{r} < 0$ with an ascending spatial flux $\phi_{c_z} > 0$ leads to a physical scenario where FMSE anomaly produced at large mesoscale structures in the near-surface region ($\xi > 0$) is overall transported toward smaller scale structures ($\phi_{r} < 0$) located above in the boundary layer ($\phi_{c_z} > 0$). Note that an overall view of the field of fluxes in the entire compound space $(r, z_c)$ (Figure \ref{fig:path}a, b and c) highlights that the near-surface and boundary layers in the mesoscale range below the cluster size are the only regions characterized by a forward cascade of energy, as also emphasized by the two vertical gray lines in Figure \ref{fig:path_average}b. The presence of such forward cascade in the boundary layer may be related to the generation of relatively small scale structures such as cold pools under the deep-convective cluster.

A different scenario appears by moving at the cluster scales of the near-surface and boundary layers (Figure \ref{fig:path}c). In particular, the forward energy cascade in the boundary layer end up at about $r \approx 280$ km (second vertical gray line in Figure \ref{fig:path_average}b), the size representative of deep convective cluster radius. Above such cross-over scale, a reverse cascade $\phi_{r} < 0$ starts to develop from the surface, opposing the forward cascade generated by the mesoscale structures. After a transition range of scales in which both forward and reverse cascades are present inside the boundary layer, only a flux toward larger scales survives for $r>300$ km. Overall, we found that the FMSE anomaly produced at the cluster scales of the near-surface region ($\xi > 0$) is transported towards even larger scale structures ($\phi_{r} > 0$) located above in the boundary layer ($\phi_{c_z} > 0$).


\subsection{Scale-space fluxes in the boundary layer top and shallow cloud layer}

The boundary layer top is the site of the most intense production phenomena of FMSE anomaly ($\xi = \xi_{max}$ at about 900 m). As shown in Figure \ref{fig:path}, this layer is also the site of the bending of the spatially ascending fluxes ($\phi_{c_z} > 0$) towards larger scales ($\phi_{r} > 0$). The strong reverse energy cascade taking origin from the boundary layer top at around 1000 m, is realized at all scales and is eventually maintained at higher atmospheric levels up to the mid troposphere. The shallow cloud layer, intermediate between the boundary layer top and the troposphere, is traversed by these spatially ascending reverse fluxes from small to large scales. This layer is a sink region in nature ($\xi <0$), suggesting that FMSE anomalies there cannot be locally sustained but rather are provided from below by the field of fluxes. Hence, the large scale structure of FMSE anomalies that realizes in the shallow cloud layer can be reasonably understood as the result of convection phenomena emerging from smaller scale structures of the boundary layer, as unequivocally shown by the field of fluxes in quantitative terms. Such reverse cascade may be related to interaction phenomena between relatively small scale structures such as cold pools that give rise to expansion or clustering phenomena above the boundary layer.  Accordingly, we conjecture that the reverse flow of scale energy from small to large scale in the shallow cloud layer is the key fundamental mechanism for the cluster generation. This finding resembles results obtained by \citet{nowak2025scale}, where they found a positive interscale transfer term at the subcloud level even if at smaller horizontal scales of about $r\approx 300$ m. Moreover, the presence of inverse cascades in more simplified framework (Rayleigh-Benard turbulent convection) have been found also in direct numerical simulations by \citet{togni2015physical} and \citet{vieweg2022inverse}.

\subsection{Scale-space fluxes in the troposphere}

The spatially ascending reverse energy cascade of scale energy variance, by entering the free troposphere, interact first with the sourcing mechanisms ($\xi >0$) of the lower troposphere between 2000 m and 3000 m, see again Figure \ref{fig:path}. These production mechanisms are active at all scales. Hence, the field of fluxes traversing the lower troposphere is fed by the local production phenomena rather than being the source of the scale energy variance of this region. In other words, the structure of FMSE anomaly of the lower troposphere is essentially the result of local but scale-wise distributed production mechanisms. 

By overcoming the lower troposphere, the spatially ascending reverse energy fluxes enter the mid troposphere (between 3000 m and 8000 m). This is a sink region for the scale-energy variance ($\xi <0$), which means that the FMSE anomalies of this atmospheric layer are essentially determined by the field of fluxes rather than by local production mechanisms. These fluxes remain first ascending ($\phi_{c_z} > 0$) and towards larger scales ($\phi_r > 0$), thus highlighting that convection from the lower troposphere generates wider structures of FMSE anomaly, see again Figure \ref{fig:path}. Accordingly, we conjecture that the reverse energy cascade in the mid troposphere is the key phenomena triggering aggregation in deep convection.

By further increasing the atmospheric elevation, the fluxes bend in the space of scales and while ascending ($\phi_{c_z} > 0$) recover a forward energy cascade towards smaller scales ($\phi_r < 0$). The switch from the reverse to the forward cascade occurs at atmospheric levels that decreases with scales, e.g. the reverse cascade penetrates the upper levels of the atmosphere more at small scales than at large scales. In particular we measure that the forward cascade starts to occur at about 5000 m for cluster scales $r \sim 300$ km (Figure \ref{fig:path}c) and at about 7000 m for meso- and local scales $r < 100$ km (Figure \ref{fig:path}a and b). At these atmospheric elevations, scale-energy variance is very weak (Figure \ref{fig:scale_var}), thus suggesting that these layers represent the upper edge of the mixing phenomena related to deep convection. The sink nature of these layers ($\xi >0$) together with the presence of a spatially ascending ($\phi_{c_z} > 0$) forward cascade ($\phi_r < 0$), suggest that the structure of FMSE anomaly at the upper edge of deep convection is the result of weak convection phenomena from larger scales at lower elevations.


\section{Phenomenological view of scale-energy paths}
\label{sec:phenomen}

Figures \ref{fig:path} and \ref{fig:path_average} have revealed the emergence of three different range of scales (local, meso and cluster scales) where different mechanisms are at play. The phenomenology of the energy paths occurring in these ranges is depicted in Figure \ref{fig:schematic}, where the dominant mechanisms for each range are reported. Figure \ref{fig:schematic} is intended to report not all possible situations that occur in a convective field, but the statistically dominant mechanisms that describe deep convective self-aggregation. 

\subsection{Local scales}

\begin{figure}[htbp]
    \centering
    \includegraphics[width=0.8\linewidth]{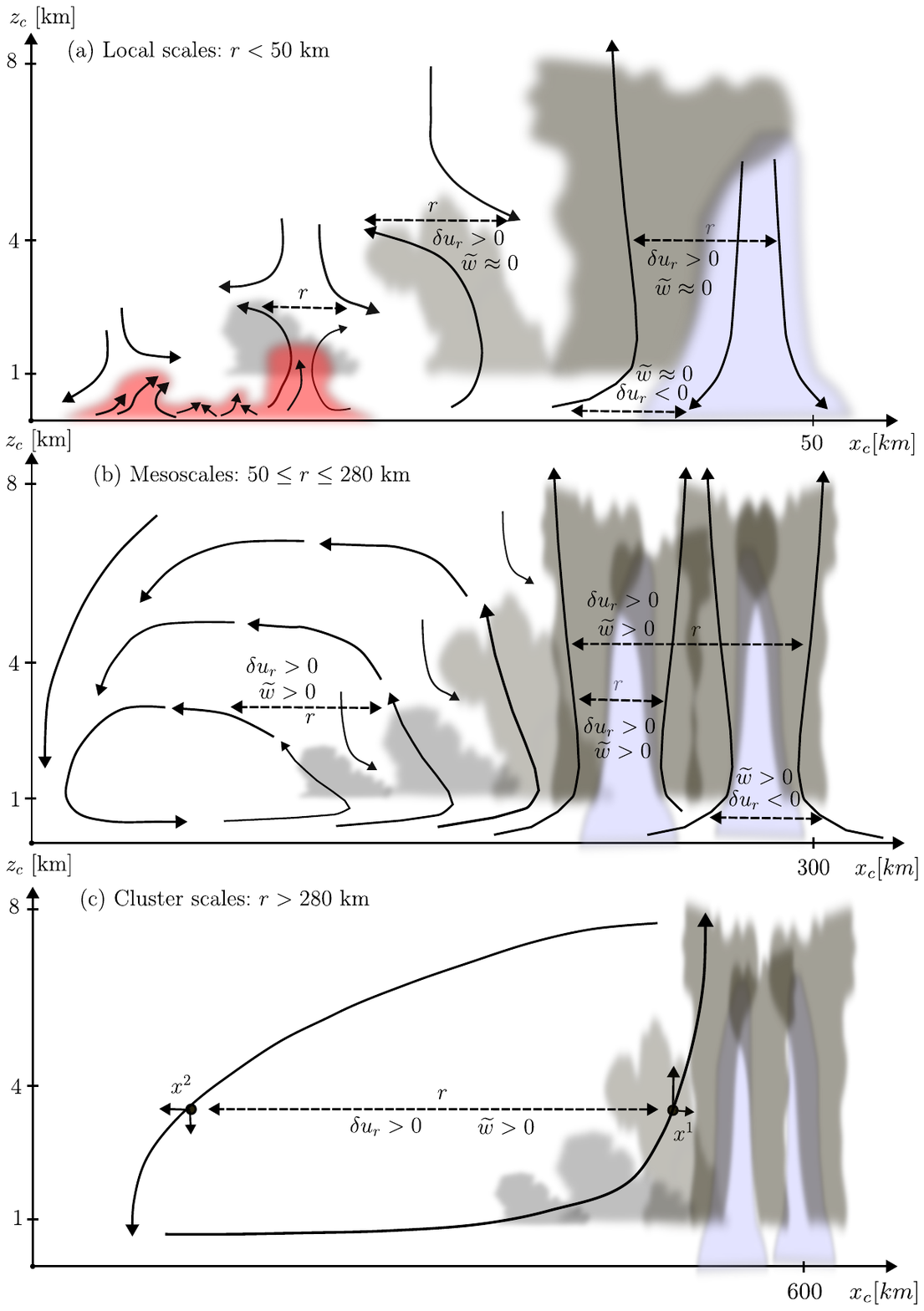}
    \caption{Phenomenological view of the scale-energy paths: a) Local scales for $r < 50$ km; b) Mesoscale for $50 \text{km} \leq r \leq 280 \text{km}$; c) Cluster scale for $r > 280$ km. Dashed horizontal lines, indicates the two-point displacement ($x^1,x^2$). Black solid arrows indicates typical streamline velocities. Gray shading indicates positive anomalies of FMSE (interpreted here as clouds). Blue shading indicates cold pool/downdraft regions, while red shading indicates plumes of positive anomalies of FMSE arising from the surface.}
    \label{fig:schematic}
\end{figure}

We summarize first the main results provided by the generalized Yaglom equation for this range (Figure \ref{fig:path}a). The range of local scales (Figure \ref{fig:schematic}a) is characterized by a small flux of energy in the physical space, $\phi_{c_z}=\rhobar \langle \delta h_f^{\prime 2}\widetilde{w'} \rangle$, which is almost constant (or decreasing) with horizontal scales below the shallow cloud layer (1500 m) up to $r \approx 100$ km (Figure \ref{fig:path_average}a). Above the boundary layer, this range of scales is characterized by a reverse cascade linearly increasing with scales. On the other hand, a smaller but forward energy cascade is observed to occur in the surface layer (Figure \ref{fig:path_average}b). Finally, the main sources/sinks of scale-energy variance (Figure \ref{fig:path_average}c and \ref{fig:path_average}d), exhibit a small increase with the horizontal scale. 

As also corroborated by the visual inspection of the instantaneous flow solutions (not shown for brevity), the small intensity of the vertical flux $\phi_{c_z}\approx 0$ can be attributed to the fact that the dominant situation in the local range of scales is a competition between convection and subsidence events. As depicted in Figure \ref{fig:schematic}a, this competition leads to $\widetilde{w} \approx 0$ on average and therefore to $\phi_{c_z}\approx 0$. This competition is stronger in the boundary layer than in the free troposphere, and is accompanied mainly by converging fluxes in the boundary and surface layer, $\delta u_r<0$, determining a small but forward energy cascade $\phi_r < 0$, and diverging/sweeping events above the boundary layer, $\delta u_r>0$, determining a reverse cascade $\phi_r > 0$. Three different mechanisms are identified for the reverse energy cascade depending on the energetic state of the region considered. In the dry regions, such sweeping events of reverse cascade $\phi_r > 0$ are created by divergence of moist air ascending from the boundary layer and divergence of dry air descending from the free troposphere, as shown by the left sketch of Figure \ref{fig:schematic}a. In the intermediate regions, it arises from the sliding layer resulting from the interaction of moist convection and dry subsidence, as shown by the middle sketch of figure \ref{fig:schematic}a. In the moistest regions, the reverse cascade $\phi_r > 0$ would result mainly by the contrast between deep convective moist updraft and downdraft, which then diverge at the surface, triggering also convergence at the surface layer and hence a forward cascade there $\phi_r < 0$ as shown by the right sketch of Figure \ref{fig:schematic}a.

\subsection{Mesoscales}

We summarize first the main results provided by the generalized Yaglom equation for this range (Figure \ref{fig:path}b). In the mesoscale range, there is a net contrast between a forward cascade of energy inside the near-surface and boundary layer ($\phi_r<0$), reaching its maximum strength at about 500 m elevation and scales about 200 km, and a strong reverse cascade above the boundary layer, reaching its maximum strength at about 1500 m and scales about 200 km. The mesoscale range is also the region of scales where the flux in the physical space, $\phi_{c_z}$, is linearly increasing with $r$ above the boundary layer (Figure \ref{fig:path_average}a), meaning that we are representing large coherent structures transporting energy from the surface upward. Also the the main components of the source term, $\Pi$ and $T_{r_z}$, as well as the scale-space transport, $\phi_r$, increase linearly with scale in the boundary layer (Figure \ref{fig:path_average}b, c and d). 
As shown in the sketch of Figure \ref{fig:schematic}b, considering horizontal scales $r > 50$ km, the dual dynamics observed in the small scale range (updraft and downdraft) is left and replaced by coherent shallow mesoscale circulations (in the intermediate regions) and larger convective systems (in the moistest regions). As shown by the right sketch of Figure \ref{fig:path_average}b, the atmospheric system at mesoscales can be considered to be composed of self-similar structures, where air is converging below the boundary layer $\delta u_r<0$ leading on average to a forward cascade $\phi_r<0$ and then diverging above the boundary layer $\delta u_r>0$ leading on average to a reverse cascade $\phi_r>0$. Considering the moist region, scales of the order of $r \approx 100$ km represent the scales of the merging of multiple cold pools and surrounding lines of convergence at the surface \cite[see Figure 1d in part I of the present work][]{silvestri2026gye_part1}. Instead, scales of the order of the cluster width, $r \approx 250$ km are representative of multiple cold pools that tend to collect again together and to form a super cold-pool and a convergence edge around the moist patch, as also described in \cite{windmiller2019convection}. Also the border dynamics of the cluster in the intermediate regions (left sketch of Figure \ref{fig:schematic}b) can be described as a self-similar structure by considering the shallow mesoscale circulations as concentric circulations which deepen and enlarge by moving toward the moistest region. In this case, a moist convective region is correlated with the outflow of another shallower convective region (see Figure \ref{fig:schematic}b), resulting in the divergence of upward fluxes, determining a reverse cascade.

\subsection{Cluster scales}

When considering the cluster as a whole and looking at scales larger than the cluster itself ($r > 280$ km), the dominating dynamics will be the large-scale circulation between the moist patch and the dry regions as sketched in Figure \ref{fig:schematic}c. In this case, the inverse energy cascade can be ascribed to two different phenomena depending on the elevation considered. Below the boundary layer, the reverse cascade is the result of an accelerating flow toward the moist region that leads to $\delta u_r > 0$ and, hence, to $\phi_r>0$ on average. Above the boundary layer, the reverse cascade is the result of the contrast between a fast ascending air in the moist region and the slow descending air in the dry region resulting again in $\delta u_r >0$ and, hence, to $\phi_r>0$ on average. Such reverse cascade, describing the expansion of the dry region and the contraction of the moist region as we move far from the surface, ends up in the upper troposphere. Indeed, already at about 7000 m, we found a slightly negative flux of energy across scales (see Figure \ref{fig:path_average}b). Finally, the contrast between fast moist convection and slow dry subsidence results also in $\widetilde{w}>0$ and, hence, on an ascending spatial flux of scale-energy variance, $\phi_{c_z}>0$ on average.

\section{Scale-by-scale budgets and reduced-order equations}
\label{sec:scale_budget}

From the analysis of the scale-energy fluxes and of the scale-energy variance of the previous sections, we have identified five different regions where different source/sink mechanisms as well as reverse/forward cascades of energy are acting: the near surface layer up to 250 m, where many terms are contributing to the production of scale energy and both reverse and forward cascade are present; the boundary layer top, where the strongest production is found and a reverse cascade is found at all scales, except for the very small ones (order of grid size); the shallow clouds layer where a dissipative region is present and the strongest reverse energy cascade processes occurs; the lower troposphere which represents a secondary source region for energy-scale variance; the upper troposphere where energy is dissipated again and only a vertical flux of scale energy survives because of the weakness of the fluxes in scale space. In this section,  we separately analyze each region considering the full form of the generalized Yaglom equation (\ref{eq:gye_RCE}) (or equivalently in a symbolic form Equation (\ref{eq:gye_symb})). From one side, this approach allows us to have a single quantitative view of the processes characterizing each atmospheric layer. From the other side, this approach allows us to address the full transport terms, including those related to the divergence in the space of vertical scales, rather than their fluxes in the space of horizontal scales. 

\subsection{Near-Surface layer and boundary layer}

\begin{figure}[htbp]
    \centering
    \includegraphics[width=1\linewidth]{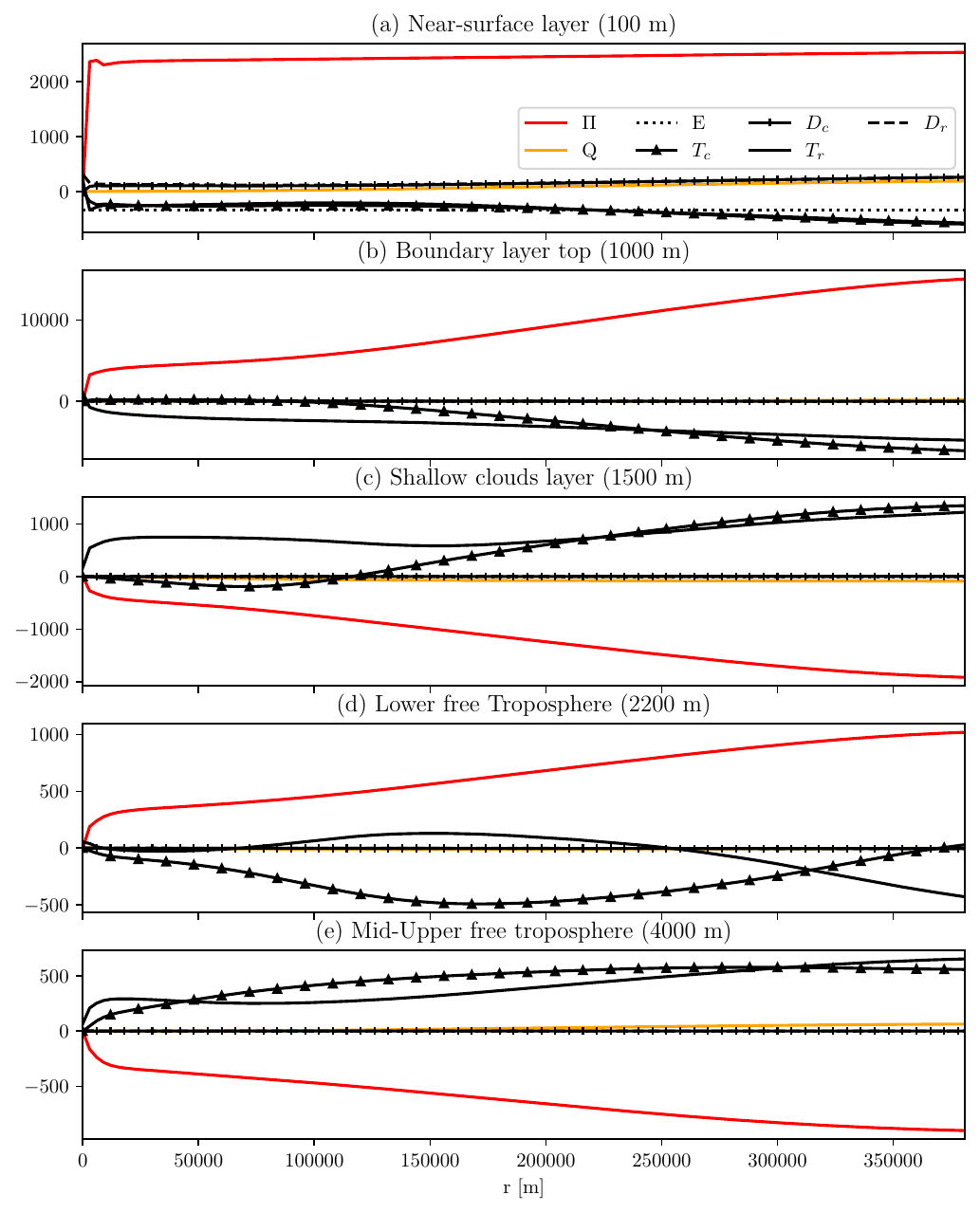}
    \caption{Scale-by-scale budget of scale-energy variance (Equation \ref{eq:gye_symb}) for different $z_c$-regions: a) near-surface layer, b) boundary layer top, c) shallow clouds layer, d) lower free troposphere, e) upper free troposphere.}
    \label{fig:scale_budg}
\end{figure}

The near-surface layer extends from the surface up to 250 m. Figure \ref{fig:scale_budg}a shows the scale-by-scale budget at about 100 m above the surface where the gradient term, $\Pi$, is a positive source and the predominant term of the budget. The gradient in this layer is caused by the heat fluxes at the surface, and it is found to rapidly increase with scales and to reach an almost constant value already at the local scale of order 10 km. Radiative heating, $Q$, also contributes as an energy source, but it is relevant only at scales larger than 50 km. Diffusive terms in physical and scale space are almost coincident and positive, meaning that sub-grid scale motions contribute to the transport of larger resolved motion in both physical, $D_c>0$, and scale space, $D_r>0$. Since spatial fluxes are always ascending (Figure \ref{fig:path}), this energy comes from the layers closest to the surface. Regarding the diffusive scale-space transport, it is important to note that the only component relevant to $D_r$ is $D_{r_z}$. Hence, the subgrid motion significantly contributes to a forward cascade in the space of vertical scale $r_z$. Also the inertial transport terms in the physical and scale space coincide, $T_c \approx T_r$, but they are negative meaning that they are absorbing energy to release it above and at other scales. The main contribution to $T_r$ is $T_{r_z}$. Hence, the only relevant transport in scale space is that related to vertical scales. In particular, the negative sign of the transport, $T_{r_z}$, suggests a draining of scale energy to be released at non zero vertical scales, $r_z \neq 0$, thus indicating a reverse cascade also among vertical scales, i.e. a process of thickening of the scale-energy structures. Overall, for the near surface layer we can write:
\begin{equation}
    D_c\approx D_r \approx D_{r_z}, \qquad T_c \approx T_r \approx T_{r_z}
\end{equation}
and the scale-by-scale budget reduces to the following form:
\begin{equation}
    -2\overline{\rho}\langle \delta w' \delta h'_f \rangle \Big( \widetilde{\frac{dH}{dz}}\Big) +2\overline{\rho}c_{pd}\langle \delta h'_f \delta Q'_{rad}\rangle - 4 \langle \widetilde{\epsilon^h_{sgs}} \rangle -2 \pdv{r_z}\langle \delta h'^2_f \delta w' \rangle - 8 \pdv{r_z}\langle\widetilde{\tau'_{h_f,z}}\delta h'_f\rangle \approx 0
\end{equation}
where also sub-grid scale dissipation $E$ must be considered because of the same order of the other terms.
Overall, the near-surface layer is a source region of scale energy variance provided by energy gradient, $\Pi$ and radiation, $Q$. This source is only partially drained by local dissipation mechanisms, $E$, and the resulting excess of scale-energy variance is drained by spatial transport $T_c$ and scale transport $T_{r_z}$ mechanisms to feed taller structures at higher elevations. The budget for the mixed boundary layer region up to 750 m is very similar to the near-surface layer and, hence, it is not shown for brevity reasons.

\subsection{Boundary layer top}

The boundary layer top is a source layer located where the production is maximum (about 900 m), just below the local maxima of the scale-energy variance (1150 m), where the gradient $dH/dz$ is large and the well-mixed boundary layer reaches its end. Figure \ref{fig:scale_budg}b is a representative picture of the layer between 750 m and 1200 m. In this region, the gradient term $\Pi$, which is the main relevant source of scale energy, is largely increasing with horizontal scales. Sub-grid scale diffusion, dissipation and radiation terms are negligible. 
For small scales, $r<100$ km, the spatial transport is also negligible, $T_c \approx 0$. At these scales, the source of energy provided by $\Pi$ is drained by the scale transport, $T_r$, which is again dominated by $T_{r_z}$. Hence, energy produced at small horizontal scales is drained to feed taller structures in the vertical direction. Therefore, the scale-by-scale budget in the small scale range of the boundary layer top reduces to the following form:
\begin{equation}
    -2\overline{\rho}\langle \delta w' \delta h'_f \rangle \Big( \widetilde{\frac{dH}{dz}}\Big) - \pdv{r_z}\langle \delta h'^2_f \delta w' \rangle \approx 0
\end{equation}
thus highlighting in a quantitative way that the small scales of the boundary layer top are dominated by a source of scale energy provided locally by the gradient term $\Pi$ that is drained by the scale-space flux $T_{r_z}$ to generate taller and, by considering the direction of the fluxes in the space of horizontal scales reported in the previous sections also wider structures through a reverse cascade. 

For scales larger than 100 km, scale energy starts to be drained also by the transport term, $T_c<0$, to feed larger distances from the surface and the budget becomes:
\begin{equation}
    -2\overline{\rho}\langle \delta w' \delta h'_f \rangle \Big( \widetilde{\frac{dH}{dz}}\Big) - \pdv{r_z}\langle \delta h'^2_f \delta w' \rangle -\pdv{z_c}\overline{\rho}\langle \delta h'^2_f \widetilde{w}\rangle  \approx 0
\end{equation}
statistically representing the ascending buoyant thermals and descending entrainment flows typical of this layer. Overall, the boundary layer top together with the near surface layer are the engine of the FMSE anomaly of the entire atmosphere, being the sit of the most intense sourcing mechanisms of scale energy that fed, through spatial fluxes and cascades, taller and wider energetic structures in the upper atmospheric layers.

\subsection{Shallow clouds layer}

Entering the dissipative layer of shallow clouds, the gradient term becomes a sink of energy $\Pi < 0$ with an increasing magnitude with scale. Sub-grid scale diffusion and dissipation are negligible $D_c \approx D_r \approx E \approx 0$, while radiative cooling $Q$ acts as a small sink of energy especially at large scales. However, this radiative effect can be neglected when compared to the sink associated to the gradient term $Q \approx 0$. Also the transport terms $T_r$ and $T_c$ are reversed with respect to the boundary layer top region. Up to scales of about 100 km there is a negligible transport in the physical space $T_c \approx 0$, while the scale space transport balances the gradient term $T_r + \Pi \approx 0$. Again, the main component of the scale space transport term is that related to a cascade among vertical scales $T_r \approx T_{r_z}$. Contrary to the boundary layer top region, the positive sign of $T_{r_z}$ denotes a source of scale energy variance from $r_z \neq 0$ scales and, hence a forward cascade among vertical scales, i.e. a thinning of energy structures in the vertical direction. At larger meso and cluster scales, $r>100 $ km, the transport among different heights $T_c$ becomes important and acts as a source. Hence, energy drained below in the boundary layer region is released here in the shallow clouds layer. Overall, the reduced budget in the shallow clouds layer has the same form of the boundary layer top region
\begin{equation}
    -2\overline{\rho}\langle \delta w' \delta h'_f \rangle \Big( \widetilde{\frac{dH}{dz}}\Big) - \pdv{r_z}\langle \delta h'^2_f \delta w' \rangle -\pdv{z_c}\overline{\rho}\langle \delta h'^2_f \widetilde{w}\rangle  \approx 0
\end{equation}
but with reversed sign for each term, thus showing the opposite role played by the underlying physical mechanism. Overall, the shallow cloud layer is a sink layer dissipating energy provided from below through spatial fluxes and from taller but, by considering the direction of the fluxes in the space of horizontal scales reported in the previous sections, also narrower scale structures through a forward cascade among vertical scales and a reverse cascade among horizontal scales. Hence, the structure of the FMSE anomalies of this region is the result of a inertial thinning and widening of energy structures provided from below by convection. In this regard, shallow clouds are the footprint of small-scale ($r<100$ km) shallow circulations damped by the stratification of FMSE found in the dissipation layer above the the boundary layer top. When the circulations are large enough ($r>100$ km), they can penetrate the dissipative layer and translate the system from a shallow to a deep convective regime, as found in previous LES studies \citep{Bretherton2017mesoscaleagg, janssens2023nonprecipitating}.

\subsection{Lower free troposphere}

Above the dissipative layer of shallow clouds, another region where the gradient term acts as a source $\Pi > 0$ is found. Again, radiation and dissipation are negligible, $Q \approx 0$ and $E \approx 0$. In this layer, the transport in scale space $T_r$ is negligible until the cluster scale is reached for $r>250 km$. Hence, at small and mesoscale, a balance between spatial transport and energy gradient source is found:
\begin{equation}
    -2\overline{\rho}\langle \delta w' \delta h'_f \rangle \Big( \widetilde{\frac{dH}{dz}}\Big) -\pdv{z_c}\overline{\rho}\langle \delta h'^2_f \widetilde{w}\rangle  \approx 0.
\end{equation}
In this case, the transport in the physical space is absorbing the scale energy variance excess provided by $\Pi >0$ to release it to the upper free troposphere $T_c < 0$. When the cluster scale is reached, then part of the energy is absorbed also by the scale space transfer term $T_r<0$, which reaches its maximum at the largest scale. Again, the scale transport is dominated by $T_{r_z}$ and the associated energy draining can be again understood to feed taller structures via reverse cascade phenomena in the vertical scales. Hence, the reduced form of the budget at cluster scales reads
\begin{equation}
    -2\overline{\rho}\langle \delta w' \delta h'_f \rangle \Big( \widetilde{\frac{dH}{dz}}\Big) - \pdv{r_z}\langle \delta h'^2_f \delta w' \rangle -\pdv{z_c}\overline{\rho}\langle \delta h'^2_f \widetilde{w}\rangle  \approx 0.
\end{equation}
Overall, the lower free troposphere is an intermediate source region of energy that sustains the ascending flux coming from the boundary layer and a reverse cascade among vertical and, by considering the direction of the fluxes in the space of horizontal scales reported in the previous sections, horizontal scales, i.e. taller and wider structures of the mid and upper troposphere are fed by this intermediate layer.

\subsection{Mid and upper free troposphere}

In the middle and in the upper free troposphere (Figure \ref{fig:scale_budg}e is representative of the layer between 3500 to 8000 m), the gradient term acts as sink $\Pi <0$, and scale energy is provided by both transport terms $T_c>0$ and $T_r>0$ at all scales. Also the radiative forcing in this case contributes positively by increasing the scale energy variance at large scales $Q>0$, but this contribution remains negligible with respect to the transport terms. Hence, a reduced form of the scale-by-scale budget for the mid and upper troposphere is the following:
\begin{equation}
    -2\overline{\rho}\langle \delta w' \delta h'_f \rangle \Big( \widetilde{\frac{dH}{dz}}\Big) - \pdv{r_z}\langle \delta h'^2_f \delta w' \rangle -\pdv{z_c}\overline{\rho}\langle \delta h'^2_f \widetilde{w}\rangle  \approx 0.
\end{equation}
Overall, the mid and upper free troposphere is a sink region dissipating energy provided from below through spatial fluxes. By considering the direction of the fluxes in the space of horizontal scales reported in the previous sections, a reverse cascade in the horizontal scales simultaneously occurs with a forward cascade in the vertical scales. Hence, the structure of FMSE anomaly of the mid and upper troposphere is the result of an inertial tinning and widening of energy structures raising from below through convection.

\section{Self-similar scaling and correlation coefficient for moisture-convection feedbak}
\label{sec:self-similar-scaling}

\begin{figure}[t]
    \centering
    \includegraphics[width=0.8\linewidth]{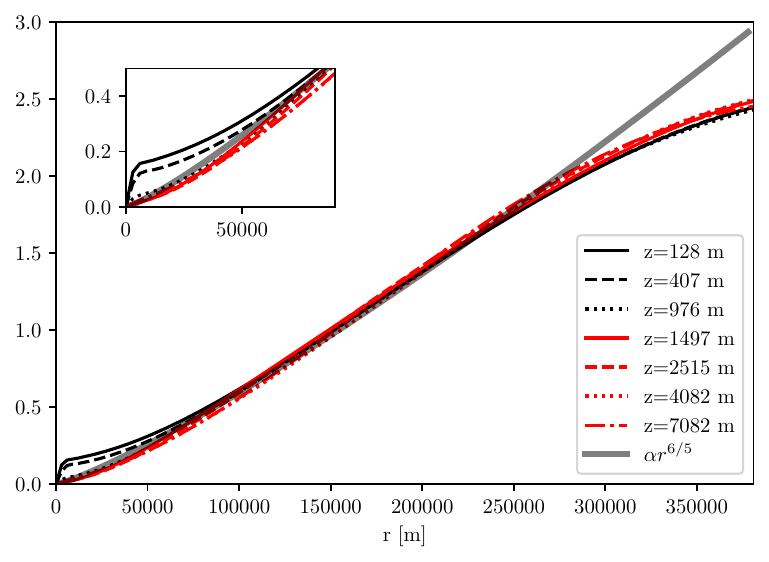}
    \caption{Scaling in the space of horizontal scales $r$ of the rescaled scale-energy variance $\langle \delta h_f'^2 \rangle/\langle h_f'^2 \rangle$ for different atmospheric levels. The thick gray line indicates the power law $\langle \delta h_f'^2 \rangle/\langle h_f'^2 \rangle = \alpha r^{6/5}$, with $\alpha=0.6/c_p^{2}$. }
    \label{fig:scale_var_levels}
\end{figure}

In this section, we focus the analysis on the scale-space scaling of relevant quantities of the generalized Yaglom equation with the aim of deriving theoretical estimates for deep convection and self aggregation. We start by considering the scale-space scaling of scale-energy variance at different atmospheric levels. In Figure \ref{fig:scale_var_levels}, the scale-energy variance rescaled with the energy variance, $\langle \delta h_f'^2 \rangle/\langle h_f'^2 \rangle$, is reported as a function of the horizontal scale $r$ and for different atmospheric levels. A well-defined scaling $\langle \delta h_f'^2 \rangle= r^{6/5}$ is observed for all levels in the mesoscale range of scales between 100 km and 250 km, indicating a self-similar behavior of frozen moist static energy variance in this range of scales, as already anticipated in Section \ref{sec:phenomen}. At cluster scales, higher than 250 km, this scaling is no more fulfilled at all atmospheric levels. On the other hand, at the smaller local scales, below 100 km, such scaling does not hold only for the boundary layer below elevations about 750 m. This feature further supports the idea that the boundary layer top represents a dividing layer. In particular, the strong stratification of the boundary layer top acts as a filter for the small scale interactions between lower and higher atmospheric levels. On the contrary, the large-scale interactions between the boundary layer and the troposphere are able to contrast the strong stratification of the boundary layer top as demonstrated by the presence of self-similar structures following the same $\langle \delta h_f'^2 \rangle= r^{6/5}$ scaling at the meso scales of both surface, boundary layer, shallow cloud layer and low-mid troposphere.

\begin{figure}[htbp]
    \centering
    \includegraphics[width=0.8\linewidth]{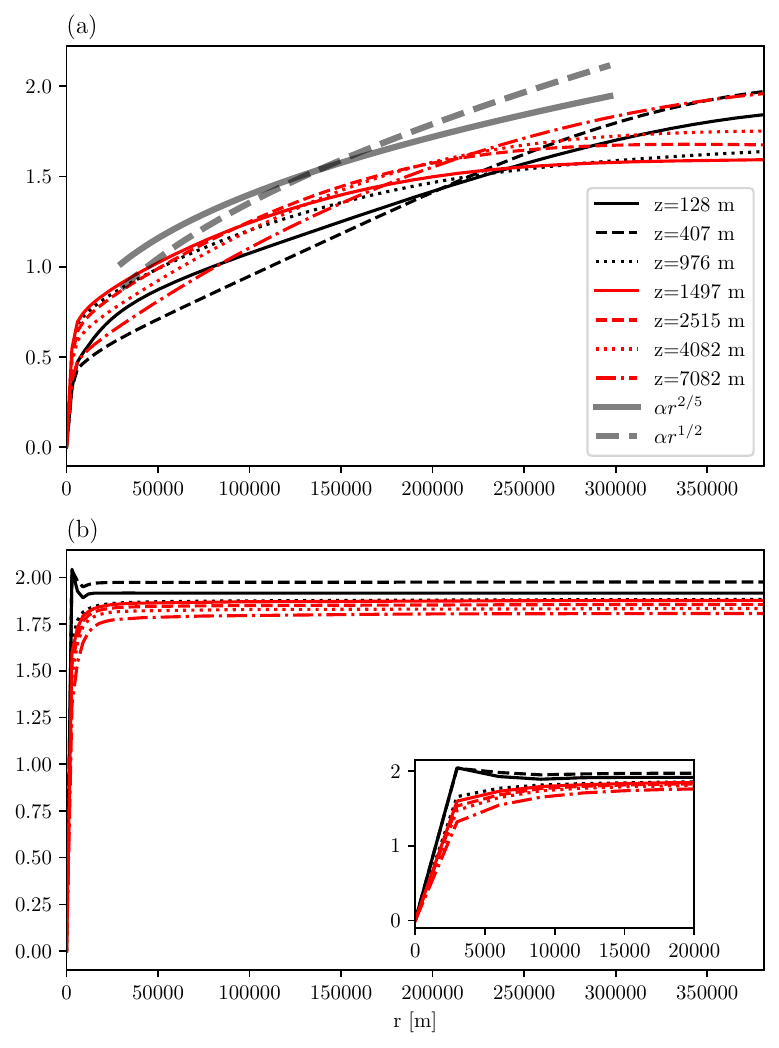}
    \caption{Scaling in the space of horizontal scales $r$ and for different atmospheric levels of (a) rescaled horizontal kinetic scale-energy variance $\langle \delta u_r'^{2} \rangle/\langle u_r'^{2}\rangle$ and of (b) rescaled vertical kinetic scale-energy variance $\langle \delta w'^{2} \rangle/\langle w'^{2}\rangle$. The thick solid gray lines indicates the power law $\langle \delta u_r'^{2} \rangle/\langle (u_r'^{2})\rangle = \alpha r^{2/5}$ while the thick dashed gray lines reports the power law($\langle \delta u_r'^{2} \rangle/\langle (u_r'^{2})\rangle = \alpha r^{1/2}$).}
    \label{fig:scaledu2_var_levels}
\end{figure}

Figure \ref{fig:scaledu2_var_levels}a and \ref{fig:scaledu2_var_levels}b show the horizontal and vertical components of kinetic scale-energy variance across different horizontal scales, respectively. Regarding the horizontal velocity increments, $ \langle \delta u_r'^{2} \rangle$, Figure \ref{fig:scaledu2_var_levels}a shows an increasing scale-energy with $r$, without a clear collapse between different atmospheric levels. At mesoscale, this increase follows two power laws with an exponent of the order of $1/2$ and $2/5$ depending on the distance from the surface: faster increase ($r^{1/2}$) for the large-scale inflow (near-surface) and outflow region (upper region); slower increase ($r^{2/5}$) from the boundary layer top region to the mid troposphere. The behavior at larger cluster scales seems to be correlated with the different slopes of increasing energy in the mesoscale range: from the boundary layer top (976 m) to the lower and mid-troposphere (4000 m) it shows a constant values with scale, while in the surface and in the boundary layer region, as well as in the upper troposphere (where the variance shows two local maxima coinciding with the large-scale inflow and outflow from the cluster), it keeps increasing with scale. 
As already shown in Figure \ref{fig:scale_var_du2}b, the vertical kinetic scale-energy variance is completely saturated at all levels for $r>30 km$. Below such scale, as shown by the inset plot in Figure \ref{fig:scaledu2_var_levels}b, the vertical kinetic scale-energy increases with scales showing also a peak value at $r=2.5$ km at vertical locations corresponding to near-surface layer and boundary layer.

We can now summarize the obtained scaling in the mesoscale range (100 km $< r <$ 250 km) as follows:
\begin{align}
    \langle \delta h_f^{\prime 2} \rangle &\sim r^{6/5}\label{eq:dh2_scale} \\ 
    \langle \delta u_r^{\prime 2} \rangle &\sim r^{2/5} \label{eq:du2_scale}\\
    \langle \delta w^{\prime 2} \rangle &\sim const \label{eq:dw2_scale}
\end{align}
The information provided by these scaling can be coupled with other important scaling obtained from the analysis of the generalized Yaglom equation and provided in Figure \ref{fig:path_average}:
\begin{align}
    \langle \delta h_f^{\prime} \delta w'  \rangle &\sim r \\
    \langle \delta h_f^{\prime 2} \delta u_r \rangle &\sim r
\end{align}
that are the linear scaling of the scale-energy variance production and the energy transport across scales in such mesoscale range. Specifically, the inverse energy cascade increases linearly reaching a maximum right at the upper limit of the mesoscale range, while production increases monotonically up to the largest scale. We can now rewrite the production and scale transport terms as the product of the statistical scaling of each variable increments and a correlation coefficient depending on $r$, $\rho_{mc} = \rho_{mc}(r)$ :
\begin{align}
    \langle \delta h_f^{\prime} \delta w' \rangle &= \rho_{mc \perp} \langle \delta h_f^{\prime} \rangle \langle |\delta w' |\rangle \sim r \label{eq:prod_scaling}\\
    \langle \delta h_f^{\prime 2} \delta u_r \rangle &= \rho_{mc ||} \langle \delta h_f^{\prime 2} \rangle \langle |\delta u_r| \rangle \sim r
\end{align}
The correlation coefficient $\rho_{mc \perp} (r)$ is a measure of the moisture-convection feedback \citep{Mapes1993gregarious, grabowski2004moisture, holloway2010temporal} since it quantifies the degree of correlation of horizontal increments of vertical velocity $\langle |\delta w' |\rangle$ and of moist static energy $\langle \delta h_f^{\prime} \rangle$. Substituting the scalings (\ref{eq:dh2_scale}) and (\ref{eq:dw2_scale}) into (\ref{eq:prod_scaling}), we obtain that this correlation coefficient must increase with scale in order to have a linear increase in production, that is:
\begin{equation}
    \rho_{mc \perp}(r) \sim r^{2/5}
\end{equation}
By repeating the same reasoning for the correlation coefficient $\rho_{mc ||} (r)$ between horizontal increments of horizontal velocity $\langle |\delta u_r'|\rangle $ and of frozen moist static energy variance $\langle \delta h_f^{\prime 2} \rangle$, we obtain:
\begin{equation}
    \rho_{mc ||}(r) \sim r^{-2/5}
\end{equation}
that is exactly the inverse rate of correlation with respect to $\rho_{mc \perp}$, i.e. the correlation decrease with scale.

\section{Theoretical estimates for self-aggregation}
\label{sec:self_aggreg_theory}

The increases of correlation with scales exhibited by $\rho_{mc \perp} (r)$ is not intuitive for classical turbulent flows. The presence of this effect in deep moist convection and the existence of the moisture convection feedback, well documented in observations as the exponential increase of rainfall with CRH \citep{Bretherton2004,holloway2010temporal}, is seen here as the necessary condition to have a linear increase of scale energy variance production with scale $\langle \delta h_f^{\prime} \delta w'  \rangle\sim r$ when the convective dynamics is fixed by the large-scale energy balance, that is $\langle |\delta w| \rangle=const$. The statistical equilibrium of convection for $r \geq 30$ km is demonstrated by the statistical independence of $\delta w$ on horizontal scale in the mesoscale range. The same result has been found by averaging rainfall over different spatial areas from \cite{islam1993predictability,bongioannini2005orographic}, resulting in a constant standard deviation of rainfall after 30 km averaging window. This is the expression of quasi-equilibrium convection, in which the vertical velocity at mesoscale is settled by large-scale equilibrium dynamics. Therefore we can state that the mesoscale self-similar range is a consequence of such quasi-equilibrium, where the vertical velocity of the cluster is settled by radiative cooling trough the weak temperature gradient hypothesis. 

Therefore we could estimate the lower and upper range of mesoscales by setting the lower and upper range validity of the weak temperature gradient across scales. The lower range above which weak temperature gradient hypothesis is valid is determined by the scale over which gravity wave dynamics have comparable scale to that of convection, so that the weak temperature gradient condition is not fulfilled. Below such lower limit, that we called $r_{meso}$, the dynamic of a single cloud system is relevant and a coherent and self-similar behavior of the system is not found. Therefore we define $r_{meso}$ such that a matching of time scales is reached:
\begin{equation}
    \tau_{gw}(r_{meso}) \approx \tau^{up}_{conv} (r_{meso})
    \label{eq:r_meso}
\end{equation}
where the gravity wave time scales can be defined as $\tau_{gw}=r_{meso}/c_{gw}$, where $c_{gw}\approx NH/\pi \approx 50$ m/s is the group velocity of the first mode of gravity waves in a stratified fluid \citep{Mapes1993gregarious} (using typical values for $N=0.01$ and $H=15000$). On the other hand, the convective turnover time can be defined as $\tau^{up}_{conv}=H/w_{up}$, where $w_{up}$ can be considered equal to 15 m/s and hence $\tau_{conv} \approx 1000 s \approx 15 min$. By using Equation \ref{eq:r_meso}, the lower limit is simply found to be $r_{meso}\approx 50$ km.

We can repeat the same analysis for the upper limit of the mesoscale range, that we call $R_{meso}^{dyn}$. In this case the upper limit to the validity of weak temperature gradient hypothesis will be broken by the effect of the coriolis force. In our simulation, the coriolis force is not considered, therefore the upper limit will be settled by the domain size. However, if we consider the self-similar behavior of mesoscale circulations to be valid in the real atmosphere, than the upper limit to the region of moist air uplift would be set by the coriolis force that would break the self-similar behavior of mesoscale circulations:
\begin{equation}
    \tau^{meso}_{conv}(R_{meso}^{dyn}) \approx \tau_{rot} (R_{meso}^{dyn})
    \label{eq:R_meso_dyn0}
\end{equation}
In this case the convective mesoscale circulation turnover time can be defined as $\tau^{meso}_{conv}=R_{meso}^{dyn}/\langle |\delta u_r| \rangle$, where $\langle |\delta u_r| \rangle \approx C_{u} r^{1/5}$, considering the more suitable scaling for the mid troposphere shown in Figure \ref{fig:scaledu2_var_levels}a and reported in Equation (\ref{eq:du2_scale}). $C_u$ is a constant that if estimated by our data for a corresponding length of $r=100$ km in the mid-troposphere, results in $C_u \approx 1$ s$^{-1}$. Instead, a different estimate could be based on typical values of trade winds of the order of $5$ m s$^{-1}$ \citep{Windmiller2024innerlife} for the mesoscale length of $r=100$ km, which give a value of $C_u \approx 0.5$ s$^{-1}$. Next the characteristic time over which the mesoscale outflow will feel the effect of earth rotation can be defined as $\tau_{rot}\approx 1/f$, with $f$ the coriolis parameter varying with latitude. If we center our cluster at the equator with $f=0$, then we estimate $\tau_{rot}\approx 1/ (\beta r)$, with $\beta=df/dy=2.29x10^{-11}$ m$^{-1}$ s$^{-1}$ being the latitudinal variation of the coriolis parameter. For a smooth transition between the equator and the other latitude we can use the formulation of the coriolis factor on a beta-plane and evaluate its effect on the distance from the cluster center (at a latitude $\Phi_0$), that is $f(\Phi)=f(\Phi_0)+ \beta r/2$. Therefore we can write $\tau_{rot}\approx1/f\approx1/(f_0+\beta r/2)$, where $f_0=f(\Phi_0)$. By equating $\tau_{conv}$ and $\tau_{rot}$ and considering the latitudinal dependence of the cluster center, we can write an equation for $R_{meso}^{dyn}$ as follows:
\begin{equation}
    \frac{f_0}{C_u} \left (R_{meso}^{dyn} \right )^{4/5}+\frac{\beta}{2C_u} \left (R_{meso}^{dyn} \right )^{9/5}=1 .
    \label{eq:R_meso_dyn}
\end{equation}

\begin{figure}
    \centering
    \includegraphics[width=0.7\linewidth]{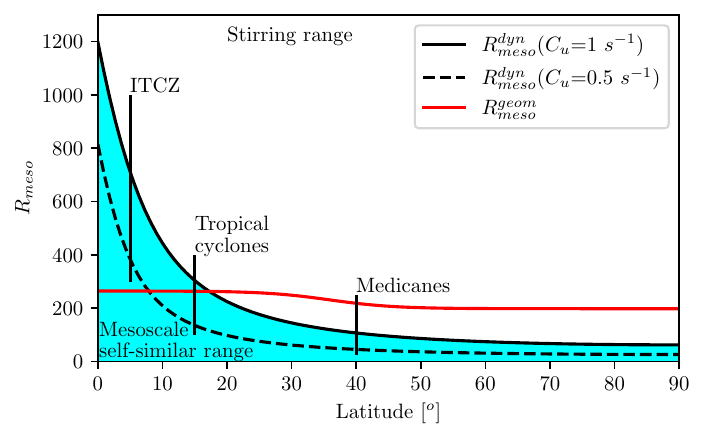}
    \caption{The critical limit of mesoscale self-similar range, $R_{meso}^{dyn}$, as a function of the latitudinal center of the cluster. Different black lines correspond to different values of the constant $C_u$, as used in the equation (\ref{eq:R_meso_dyn}). Red line indicates the geometrical limit $R_{meso}^{geom}$.}
    \label{fig:ictz_width}
\end{figure}

The numerical solution of Equation (\ref{eq:R_meso_dyn}) is reported in Figure \ref{fig:ictz_width}. In particular, Figure \ref{fig:ictz_width} shows how the natural scale of mesoscale aggregation, that is aggregation in the absence of large-scale external forcings other than surface-fluxes and radiative cooling, would change by changing the latitude of the cluster center. If we consider the location of the deep convective organization at the equator, we would obtain a width which ranges between 800 km and 1200 km (for different values of the constant $C_u$). By moving the cluster center to $5^o$ of latitude, which is representative of the ITCZ location, the maximum width of self-similar mesoscale range would be between 400 and 800 km. Such values are in agreement with typical range of ITCZ width \citep{Dias2011modulation,Windmiller2024innerlife}. Overall, increasing the latitudinal center of the convection, the effective radius over which mesoscale aggregation would operate decreases, reaching typical values of Tropical cyclones scales (200-400 km) around 15$^o$ and Medicanes (30-150 km) around 40$^o$. It is interesting to notice that the upper limit at higher latitudes correspond to the values obtained by the estimate of $r_{meso}$, that is between 30 and 60 km. Above this critical limit, the mesoscale aggregation would feel the stirring action of the coriolis force. This is why we called the range of scales above the critical limit $R_{meso}^{dyn}$ as the stirring range.

In the absence of any rotation, such as in the RCE idealized setup, the above argument would imply a continuous grow of the mesoscale system up to the scale of the domain size. However, in our simulation as well as other simulation in literature \citep{muller2012detailed,Yanase2020critical}, the cluster occupy less than half of the domain size and it has a width of the order of $280$ km, coinciding with exactly the end of the mesoscale range. Moreover, long-channel geometry simulations \citep{wing2016self,wing2020clouds} as well as larger squared domains \citep{arnold2018nonrotating,patrizio2019sensitivity}, have shown the appearance of multiple clusters that eventually merge together forming a single cluster. Multiple clusters and breaking/merging phenomena have found to occur for domains where at least one dimension is larger than O(1000 km). In any case also the characteristic size of such super-clusters shows an upper limit as demonstrated by \citet{arnold2018nonrotating} and \citet{patrizio2019sensitivity}. Here we want to introduce an upper limit to the single cluster that may develop in relatively small domains, before large-scale merging/breaking occurs due to different physical mechanisms \citep{patrizio2019sensitivity}. If we hypothesize that the self-similar grow of mesoscale circulations is fundamental for the maintenance of the cluster, in the absence of rotation, an upper limit to the maximum horizontal extension of a mesoscale circulation would be set by its maximum vertical extension. In this view, we called this limit as geometrical and indicate it as $R_{meso}^{geom}$, and the maximum height that a mesoscale circulation can reach would be set by the tropopause height, $H$. The link between the horizontal $r$ and vertical scale $r_z$ of a mesoscale circulation can be found by considering the budget of momentum in the vertical direction and the continuity equation:
\begin{align}
    \frac{1}{\rhobar} \pdv{\rhobar w'^2}{z}  \approx g \frac{h_f^{\prime}}{\overline{h}} \\
    \frac{1}{r} \pdv{r u_r'}{r} + \frac{1}{\rhobar} \pdv{\rhobar w'}{z} =0
\end{align}
where $\overline{h}=\overline{h}(z)$ is the horizontally averaged profile of $h_f$. The augmented space addressed in this work can now be used to rewrite the above equations using increment variables. Indeed, the increment formalism strictly connects with differential equations, where derivatives of a generic quantity $g$ are ratios $\delta g/r$ taken at different scales. Hence, the above equations can be rewritten in a scale-by-scale form as:
\begin{align}
    \frac{\delta_z w^{\prime 2}}{r_z} \sim g\frac{\delta h'_f}{\overline{h}} \label{eq:mom_bal}\\
    \frac{\delta u'_r}{r}+\frac{\delta_z w'}{r_z} \sim 0 \label{eq:mass_bal}
\end{align}
where the symbol $\delta_z$ is used to indicate the increment along the vertical in contrast to $\delta$, which is the increment along the radial direction (for $r_z=0$). In order to simplify the momentum equations, we have assumed that: a) weak-temperature gradient holds in the mesoscale range, that is $\delta B \approx g \delta h'_f/\overline{h}$; b) we are in the mid troposphere where we can neglect sub-grid stresses; c) that the horizontal increment of vertical velocity are constant in the mesoscale range as found before (quasi-equilibrium of convection). The assumed balance between the vertical energy flux term and the buoyancy term in the mesoscale spectrum of horizontal scales (between 30 km and 200 km) has been recently confirmed by energy spectrum analysis in the upper troposphere (between 6-10 km) by \cite{lai2023kinetic} under RCE idealized simulation of deep convective self-aggregation.

Equation (\ref{eq:mom_bal}) can be used to find a scale for the vertical velocity increment, $\delta w' \approx \sqrt{g r_z \delta h'_f/\overline{h}}$. By substituting this estimate for the vertical velocity scale into the scale-continuity equation (\ref{eq:mass_bal}) and using the scaling for $\delta h_f', \delta u_r$ found in the mesoscale range (Equations \ref{eq:dh2_scale} and \ref{eq:du2_scale}), an equation relating the horizontal and vertical scales of mesoscale circulations can be written:
\begin{equation}
    r_z \sim \frac{\delta h'_f}{\delta u_r^2}r^2 \sim r^{11/5}
    \label{eq:aniso_scaling}
\end{equation}
This equation shows that if the vertical velocity divergence balances the horizontal anomalies of moist static energy in the mesoscale range, then the vertical extension of the mesoscale circulation must increase almost following the square of the horizontal extension. Then the maximum horizontal extension, $R_{meso}^{geom}$, would be the horizontal scale that can be computed from Equation (\ref{eq:aniso_scaling}) by considering $r_z=H$. From Figure \ref{fig:path_average}a, we have seen that the vertical flux $\phi_{c_z}$ start to increase linearly with $r$ at about $H_0=$2000 m height and at a reference scale of $r_0$=100 km. By using such reference value as proportionality constants we can write:
\begin{equation}
    R_{meso}^{geom}=r_0 \Big(\frac{H}{H_0}\Big)^{5/11}
\end{equation}
In the tropical atmosphere, where $H=$17000 m, such scaling holds $R_{meso}^{geom}\approx260$ km, an estimate which is inline with the upper limit found in the numerical simulation. By assuming a tropospheric height that is varying with latitude between $H_{eq}=17$ km at the equator toward $H_{pole}=9$ km, we can draw this geometric limit as a function of latitude, as reported in Figure \ref{fig:ictz_width}. At the pole this maximum geometrical scale will decrease to $R_{meso}^{geom}\approx200$ km. It is interesting to notice that the two limits ($R_{meso}^{dyn}$ and $R_{meso}^{geom}$) intersect around 20$^o$ of latitude that is the end of the tropical region. This means that inside the tropical regions, the mesoscale self-similar limit would be set by the tropopause and other mechanisms, such as external forcings, merging, large-scale breaking, that would be responsible for larger observed aggregated cluster. On the other hand, the more stringent dynamical limit $R_{meso}^{dyn}$ is valid for latitude above 20$^o$, and larger clusters should be related to other processes other than spontaneous aggregation due to internal feedbacks. 


\section{Conclusions}
\label{sec:conclusions}

This work extends to the space of scales, the analysis of idealized simulations of RCE made for the physical space in part I of the present research \citep{silvestri2026gye_part1}. The novel theoretical framework provided by the generalized Yaglom equation is here derived and used to address for the first time the multi-dimensional nature of deep convection and the processes that maintain the final aggregated state. The generalized Yaglom equation is the evolution equation for the second-order structure function of frozen moist static energy, which we refer to as scale-energy variance, $\langle\delta h_f^{\prime 2}\rangle$.  This theoretical framework allows us identify in an unequivocal way a multi-dimensional field of fluxes transporting scale-energy variance among different atmospheric levels $\phi_{c_z}$, and among different scales $\phi_r$. This field of fluxes is generated and eventually dissipated by source and sink mechanisms $\xi$, which are again quantitatively described in the compound space of scales and atmospheric elevations.

\subsection{Energy and velocity structures}

In part I of this work, we found that the moist static energy variance has a bottom-heavy vertical profile with a maximum in the boundary layer top region around 1150 m. The scale-space analysis shows that the scales carrying most of the energy variance are found to be larger than 100 km in the near surface layer and in the boundary layer region below 750 m. Such convective turbulent length scales decrease to 30 km in the boundary layer top region and then increase again up to reaching the largest domain scale at about 5000 m above ground. The relatively large length scales of the boundary layer can be related to large cold-pool structures that organize convection in larger arcs of length of the order of 100 km. On the other hand, the small scales of the boundary layer top region, between 30 km and 100 km, appear to be critical in the processes governing the production of scale-energy variance or equivalently of FMSE anomalies.

The second-order structure function of the velocity field reveals that the surface and boundary layers are fully mixed regions where turbulence mixing is active down to very small scales of order 5 km. On the contrary, the turbulence scales of the free troposphere are larger and effective for lengths larger than 80 km. In between these two regions, the atmospheric layer centred around 850 m and here referred to as the boundary layer top, is found to act as a sliding and deformable atmospheric layer that separates the well mixed turbulent dynamics of the lower atmospheric levels from the larger convective structures of the troposphere. Of relevance, it is also found that the vertical velocity structure function rapidly saturates to a constant value for scales above 30 km, thus demonstrating that quasi-equilibrium of convection with large-scale dynamics is already found at such small scales, in agreement with previous studies \citep{islam1993predictability, bongioannini2005orographic, yano2025emergence}.

\subsection{Dynamics of scale-energy variance from its source to its sink}

The analysis of the source and sink regions of scale-energy variance confirms the layered structures of the atmosphere reported in part I of the present work \citep{silvestri2026gye_part1}. The surface and boundary layers are found to be the sit of the most intense production mechanisms of FMSE anomaly $\xi > 0$ with a maximum $\xi_{max}$ located at the boundary layer top around 900 m (right below the peak of scale-energy variance). These sourcing phenomena are found to be related to the mean gradient of FMSE, $\Pi = -2 \rhobar  \langle \delta h_f' \delta w' \rangle (dH/dz)$ and are found to be active from large to small scales of the order of 10 km, again supporting the idea that these layers are fully mixed regions where small-scale structures also participate in the production of FMSE anomalies. Above the boundary layer top, a layered structure of alternating sink and sources is found as determined by the sign of the mean gradient of FMSE through the term $\Pi = -2 \rhobar  \langle \delta h_f' \delta w' \rangle (dH/dz)$: a sink region in the shallow cloud layer $\xi < 0$; a weak secondary source region in the lower free troposphere $\xi > 0$ and again a sink region in the mid-upper free troposphere above 3000 m up to 8000 m $\xi < 0$. It is relevant to note that the structure of FMSE anomaly of the sink regions is essentially determined by the scale-energy variance provided by the field of fluxes, contrary to the source regions where local production mechanisms actively provide scale-energy variance. Accordingly, the structure of FMSE anomaly can be inferred by addressing the production mechanisms in the source regions (surface and boundary layers) while the field of fluxes in the sink regions (shallow cloud layer and mid-upper free troposphere).

The multidimensional field of fluxes revealed that all scales and all layers are traversed by an ascending spatial flux ($\phi_{c_z}>0$) that drains scale-energy from the surface and boundary layers and releases it in the shallow cloud layer and in the mid-upper troposphere. Instead, a more complex behavior is found for the flux in the space of scales ($\phi_r$). In the surface and boundary layers, the scale-energy variance produced in the mesoscale range, $100 \lesssim r \lesssim 250$ km, is found to transferred to feed FMSE anomalies at higher atmospheric levels and narrower but taller scales (ascending direct cascade among horizontal scales and reverse cascade among vertical scales) while the scale energy variance produced in the cluster range, $r > 280$ km, is found to be transported to generate FMSE anomalies at higher elevations and wider and taller scales (ascending reverse cascade among both horizontal and vertical scales). By entering the boundary layer top, these fluxes are further energized by the intense production mechanisms of this layer but are also all bent towards larger horizontal scales. Hence, only an ascending reverse cascade is found to survive thus supporting again the idea of the boundary layer top as a buffer layer dividing the large scale circulation dynamics of the free troposphere from the fully mixed scale-wise phenomena of the lower atmospheric levels. 

Overall, the lower atmospheric levels up to the boundary layer top are found to be the engine of FMSE anomaly of the entire atmosphere, being the sit of the most intense sourcing mechanisms of scale energy that fed, through spatial fluxes and cascades, taller and wider energy structures in the upper atmospheric layers. Indeed, the shallow cloud layer and the mid-upper troposphere are sink regions and, hence, the FMSE anomaly there can be only sustained by fluxes from other regions. Interestingly, these transport mechanisms invert to a forward cascade in the space of vertical scales by entering these regions. In particular, the tall and wide structures emerging from the boundary layer top through convection are found to experience a thinning but still widening action in the shallow cloud layer (ascending forward cascade in the vertical scales and reverse cascade in the horizontal scales). On the other hand, the energy structures emerging from the shallow cloud layers and energized by the source mechanisms of the lower troposphere, are found to experience first a thinning but widening action in the mid troposphere (ascending forward cascade in the vertical scales and reverse cascade in the horizontal scales) and then a thinning and also shrinking action in the upper troposphere (ascending forward cascade in the vertical and horizontal scales).

Based on these results, the debated phenomenon of self-aggregation (transition of clouds from a random state homogeneously distributed across the domain into a more organized state with a moist convective patch surrounded by a dry patch free of convection) is found to be sustained by the reverse cascade associated with convection in the lower free troposphere, which transfers, from small to large scales, energy provided by the production mechanisms of the surface and boundary layers, through the development of shallow mesoscale circulations. This finding allows us to interpret and explain the observed scatter of results about self-aggregation reported in literature.  The related discussion is reported in Section \ref{sec:disc}\ref{subsec:disc_conditions}. In this scenario, reduced-order models for the inverse energy cascade are demanding. A new paradigm needs to be adopted in the parametrizations of climate models to represent such cascades. Indeed, as shown by \citet{bao2024intensification}, the degree of convective organization directly controls the intensification of tropical precipitation extremes, making the parameterization of this inverse cascade a pressing target for improving the representation of extreme rainfall in a warming climate.


\subsection{Atmospheric phenomena involved and the filtering role of the boundary layer top}

The forward cascade in the space of horizontal scales is found to be related to phenomena of converging flow ($\delta u_r < 0$) while the reverse cascade is found to be associated to diverging/sweeping motions ($\delta u_r > 0$); the specific mechanisms generating these motions depend on whether the region is dry, intermediate, or moist. At local scales ($r \lesssim 50 - 100$ km), the spatial flux is weak and nearly constant, consistent with a near-cancellation between upward and downward fluxes ($\widetilde{w} \approx 0$) and a near-zero forward cascade near the surface. At the mesoscales ($100 \lesssim r \lesssim 250$ km) of the surface and boundary layers, a transition is observed from the dual updraft/downdraft dynamics to self-similar, coherent structures (shallow mesoscale overturning circulations and larger convective systems) characterized by strong, concentrated ascending motion and weaker, more diffuse descending motion, as is typical of mesoscale organization. This asymmetry is found to sustain a net forward cascade within the boundary layer and a strong reverse cascade above it, both increasing linearly with scale and peaking around $r \approx 200$ km. On the other hand, at scales larger than the cluster itself ($r > 280$ km), the reverse cascade in the surface and boundary layers is found to be consistent with large-scale circulation dynamics between moist and dry regions. This phenomenon sustains a reverse cascade throughout the column ($\delta u_r > 0$, $\widetilde{w} > 0$) that progressively weakens with height and reverses sign in the upper troposphere ($z \approx 7000$ m).

The inverse cascade of energy found in the shallow cloud layer is realized through a series of shallow and mid-level mesoscale circulations which behave in a self-similar way. Indeed, an evident degree of self-similarity has been found for the convective/subsidence energy structures at the mesoscale range (100-250 km) as demonstrated by the almost perfect collapse of scale-energy variance to the power law scaling $\langle \delta h_f^{\prime 2} \rangle \sim \langle h_f'^2 \rangle r^{6/5}$. This scaling does not hold at the smaller local scales, below 100 km, only for the surface and boundary layers below elevations about 750 m. As a consequence, we found that the boundary layer top represents a scale-dependent dividing layer between low and high atmospheric regions. In particular, the stratification of the boundary layer top is found to be strong enough to filter out the interactions between lower and higher atmospheric levels at small local scales $r\lesssim 100$ km. On the contrary, the large-scale interactions between the boundary layer and the troposphere are found to be able to contrast the strong stratification of the boundary layer top as demonstrated by the presence of mesoscale self-similar structures following the same power law scaling and spanning all atmospheric layers.

\subsection{Theoretical estimates based on self-similar circulations}

The nearly perfect self-similar behaviour of scale-energy variance across heights $\langle \delta h_f^{\prime 2} \rangle \sim \langle h_f'^2 \rangle r^{6/5}$ indicates that, despite the strong vertical inhomogeneity of the convective field, the horizontal structure of moist static energy anomalies in the mesoscale range is governed by a single, height-independent scaling law.  Analogously, we found a degree of self-similarity also for the horizontal velocity increments, $\langle \delta u_r^{\prime 2} \rangle \sim \langle u_r^{\prime 2} \rangle r^{2/5}$, despite a residual height dependence is retained. On the other hand, the vertical velocity increments, $\langle \delta w^{\prime 2} \rangle$, are found to saturate to a constant value already for local scales around 30 km. These scalings have been used to quantify explicitly the moisture-convection feedback, expressed through the scale-dependent correlation coefficients $\rho_{mc\perp}(r)$ and $\rho_{mc\parallel}(r)$ linking, respectively, the vertical energy transport and the horizontal energy production to the underlying increments of $w'$, $u_r$, and $h_f'$. We found that the two correlation coefficients must scale as $\rho_{mc\perp} \sim r^{2/5}$ and $\rho_{mc\parallel} \sim r^{-2/5}$: the correlation between vertical velocity and moist static energy anomalies is found to strengthen with scale, while the correlation between horizontal convergence and moist static energy variance is found to weaken with scale, at exactly the inverse rate. 

This behavior provides a direct quantitative link between the observed energy-variance cascade and the well-documented moisture-convection feedback inferred from the exponential relation between precipitation and column relative humidity \citep{Bretherton2004,holloway2010temporal}. Based on a quasi-equilibrium convective framework ($\delta w \approx$ const), we have formulated hypotheses for the lower and upper bounds of the mesoscale self-similar range, and linked them to the natural width of the ITCZ. The lower bound, $r_{meso} \approx 50$ km, is set by requiring the weak-temperature-gradient time scale of convective plumes to match the propagation time scale of gravity waves; below this scale, individual cloud systems dominate and no coherent self-similar behavior emerges. The upper bound is instead governed by two competing constraints. A dynamical limit, $R_{meso}^{dyn}$, is derived from the requirement that the convective turnover time of the mesoscale circulation matches the time scale over which the Coriolis force becomes effective, and it therefore depends explicitly on latitude, shrinking from about 800--1200 km at the equator to typical tropical-cyclone and Medicane scales at higher latitudes. A geometrical limit, $R_{meso}^{geom}$, instead is derived from the vertical momentum and continuity equations under the relevant finding that the vertical extent of a mesoscale circulation scales with its horizontal extent as $r_z \sim r^{11/5}$, so that the maximum horizontal scale is set by the requirement $r_z = H$, the tropopause height. This yields $R_{meso}^{geom}\approx260$ km in the tropics, consistent with the cluster width found in our simulation. The two limits $R_{meso}^{dyn}$ and $R_{meso}^{geom}$ are found to intersect at about $20^\circ$ latitude, suggesting that within the tropical belt the self-similar mesoscale range --- and hence the natural aggregation scale, closely related to the ITCZ width --- is set primarily by the geometrical constraint of the tropopause height, while poleward of this latitude the more restrictive dynamical (rotational) limit takes over. 

Overall, the scale and position dependent framework developed here offers a useful paradigm for interpreting and modelling deep convective self-aggregation in the idealized RCE configuration. The paradigm is based on the postulate that self-similar mesoscale circulations represent the natural unit of self-aggregation reaching maximum width of about 250 km in the Tropics, following the here proposed theoretical estimates. In this framework, larger values of ITCZ width (typical values range from 400 km to 800 km \citep{Dias2011modulation,Windmiller2024innerlife}) should be considered as large-scale merging of multiple natural units of self-similar circulations. Further discussions on the self-aggregation limits are reported in Section \ref{sec:disc}\ref{subsec:discuss_limits}.
 
\section{Discussions}
\label{sec:disc}

\subsection{Discussion on the conditions for self-aggregation}
\label{subsec:disc_conditions}

The reverse flow of scale energy from small to large scale in the shallow cloud layer and the related production mechanisms located below in the surface and boundary layers, are found in Section \ref{sec:path_scale_var} to be the key fundamental mechanisms for the cluster generation. This finding allows us to interpret and explain the observed scatter of results about self-aggregation reported in literature. As demonstrated by \citet{muller2012detailed} and \citet{Yanase2020critical} using two different models, simulations of convection in domains of the order of 100 km do not exhibit self-aggregation, regardless of the model's horizontal grid resolution. In such small domains, the mesoscale range (where the reverse energy-variance cascade is found to be stronger) can only be partially represented, and its effect is suppressed by the physical constraint imposed by the domain size. Moreover, we have shown that the largest production of scale-energy variance, associated with the source terms $\Pi$ and $T_{r_z}$, is itself concentrated at these larger scales, growing linearly with $r$ and peaking around $r \approx 200$ km. In a domain smaller than 100 km, this production is therefore truncated before it can develop, further limiting the emergence of the coherent mesoscale circulations required for expanding the dry patches eventually created during the initial random phase of convection. Conversely, self-aggregation in a small domain is prevented by the forward cascade sustained by cold-pool dynamics in the boundary layer. The net effect of cold pools at local scales is to weaken the upward convective flux and to redistribute energy toward smaller scales, which is what prevents $\phi_{c_z}$ from increasing linearly with $r$ below 100 km. This raises the question of whether self-aggregation could nonetheless be achieved in a small domain. In principle, this may be possible by artificially reducing the scale at which cold-pool-driven suppression occurs (of the order of 30--100 km), for instance by suppressing rain evaporation and the associated cold-pool dynamics as found in \citet{jeevanjee2013convective} or modifying their overall properties by changing microphysics \citep{casallas2025sensitivity}. Another way could be by saturating the sub-cloud layer as found to spontaneously develop in ARPS model by \citet{cerlini2023competing}.

While the reverse cascade is directly responsible for the generation large-scale aggregate of FMSE anomaly, the production mechanisms at the boundary layer top that provide scale energy for the reverse cascade, can be equivalently, although indirectly, considered as responsible for self-aggregation. These production mechanisms are mainly induced by mean gradients of FMSE through the term $\Pi = -2 \rhobar  \langle \delta h_f' \delta w' \rangle (dH/dz)$ rather than radiative cooling anomalies $Q=2 \rhobar c_{pd} \langle \delta h_f' \delta Q_{rad}' \rangle$. However, as highlighted in part I of the present research \cite{silvestri2026gye_part1}, radiative cooling plays a fundamental role in determining the mean vertical gradient of FMSE $dH/dz$ and, hence it indirectly has an impact on the intensity of the production mechanisms $\Pi$ in the boundary layer top region. This result is also in line with the result from LES simulations of shallow cumulus clouds as found by \citet{Bretherton2017mesoscaleagg} and \citet{Janssens2024shallow}. Since the production mechanisms $\Pi$ in the near surface layer and in the boundary layer top region are fundamental for convective aggregation, any modification to the mean radiative cooling profile could alter the local production mechanisms and, hence, self-aggregation. As a matter of facts, \citet{jenney2023vertical} shows how self-aggregation can be triggered in SAM model in small domains (128 x 128 km) by strongly increasing the vertical resolution of the grid. Finer vertical resolution were found to increase the mean radiative cooling and generating a drier free troposphere. Following our results this may increase the local production $\Pi$ at all scales and reducing the scale at which reverse cascade in the shallow cloud layer start to be active. Moreover, due to the radiative convective equilibrium state, an increase of radiative cooling will also result in an increase of surface fluxes, and a stronger vertical spatial flux at smaller scales which could make the term $\phi_{c_z}$ increasing linearly with scale at smaller scales. The same sensitivity can be found with regards to the horizontal resolution and sub-grid scale mixing parametrization. As shown in \citet{silvestri2024numerical} and \citet{casallas2025sensitivity}, a larger sub-grid scale mixing impacts the size and distribution of convective core properties, increasing the free convective distance and rendering convection less efficient in moistening the lower and mid-free troposphere. Unless enough numerical diffusion is supplied, the lower-free troposphere will become enough dry to enhance mean radiative cooling and to sustain the production mechanism at the boundary layer top region which feed the reverse cascade. The strong sensitivity of self-aggregation to planetary boundary layer mixing schemes \citep{casallas2025sensitivity,shi2021modulation} is again a confirmation of the relevance of this region for self-aggregation. In future, applying the Generalized Yaglom Equation framework to such sensitivity studies and increasing further the limited horizontal and vertical resolution of the present study would help in better resolving local scales and vertical gradients and better understanding the local mechanisms competing in the creation/removal of frozen moist static energy anomalies.

\subsection{Discussion on critical limits for self-aggregation}
\label{subsec:discuss_limits}

In Section \ref{sec:self_aggreg_theory}, we postulate that self-similar mesoscale circulations represent the natural unit of self-aggregation reaching maximum width of about 250 km in the Tropics, following the proposed theoretical estimates. However, typical observed range of the ITCZ ranges from 400 km to 800 km \citep{Dias2011modulation,Windmiller2024innerlife} and, in the present theoretical framework, can be explained as large-scale merging of multiple units of self-similar circulations. \citet{patrizio2019sensitivity}, using larger domains, indeed found increasingly larger clusters as domain size increases. However, in such large-domain simulations the final cluster size is also controlled by merging between multiple smaller clusters that initially develop across the domain and subsequently coalesce, a mechanism that is more related to the interaction of multiple self-similar mesoscale circulations discussed here rather than to their growth.

In the developed theoretical framework, Section \ref{sec:self_aggreg_theory}, the upper limit of clusters for latitudes above 20$^\circ$ is imposed by the dynamical limit $R_{meso}^{dyn}$ given by the Coriolis force. The role of Earth's rotation in setting this upper limit is consistent with observations and simulations of planets with different rotation rates: a slower-rotating planet would be expected to exhibit a wider ITCZ, since the self-similar mesoscale range could extend further before being broken by stirring \citep{faulk2017effects}. This is in line with our estimate of $R_{meso}^{dyn}$, which increases as the Coriolis parameter decreases toward the equator. The breaking of inverse cascade processes with the introduction of Coriolis force also resembles results from direct numerical simulation of turbulent Rayleigh-Benard convection as found in \citet{vieweg2022inverse}. Concerning the role of latitude, an idealized RCE simulation on a beta-plane by \citet{Carstens2023regimes} showed a gradual transition of clustering regimes from the equator toward mid latitudes, even in the absence of differential surface forcing (i.e. without imposed SST gradients). This supports our argument that self-aggregation should exhibit a gradual regime shift with latitude, consistent with the latitudinal dependence of $R_{meso}^{dyn}$ and $R_{meso}^{geom}$ derived in Section \ref{sec:self_aggreg_theory}. It should be noted, however, that rotation may also modify the mechanisms regulating the mesoscale range itself, possibly altering the self-similar scalings (Equations \ref{eq:dh2_scale}--\ref{eq:dw2_scale}) rather than simply truncating them, a possibility that our non-rotating framework cannot address directly.

Finally, we acknowledge that the scalings and critical aggregation limits derived in Section \ref{sec:self_aggreg_theory} are likely to depend on sea surface temperature and other external boundary conditions, in addition to rotation. Nonetheless, we believe this scale-analysis framework offers a useful paradigm for interpreting deep convective self-aggregation in the idealized RCE configuration. The same approach could be generalized to numerical setup involving rotation or mean wind shear. More broadly, this framework allows us to quantify each of the physical processes involved in convective self-aggregation, both in physical space and in scale space, and, building on the scale-by-scale energy budget, to derive reduced-order models of the inverse energy cascade. A new paradigm needs to be adopted in the parametrizations of climate models to represent such cascades and to treat buoyancy-generated plumes, perhaps with embedded clouds and precipitation, differently from mechanical shear-generated eddies \citep{zilitinkevich2021order}. This is not merely a matter of dynamical completeness: as shown by \citet{bao2024intensification}, the degree of convective organization directly controls the intensification of tropical precipitation extremes, making the parameterization of this inverse cascade a pressing target for improving the representation of extreme rainfall in a warming climate.

\acknowledgments
L.S. and P.B.C was financed by the PR FESR 2021-2027 project - Action 2.4.1 “Prevention of Natural Hazards” - RIMU-CLIMA Project “Integrated Umbrian Meteorological Network and Tools for Climate Analysis in Umbria” (CUP:  I62B25000210006) Italian Ministry of University and Research (MUR). Moreover, this research has been also carried out as part of the PRIN 2022 PNRR Project P20229KW2R - SEAPLANE - "Simulation and modelling of interface fluxes in wind-wave flows for an improved climate science", CUP E53D23017010001, funded by the National Recovery and Resilience Plan (PNRR), Italy, Mission 04 Component 2 Investment 1.1 – NextGenerationEU. We thank Dr. Gabriele Boga for useful discussion on this work, especially in the phenomenological view of the energy paths.

%
%
\datastatement
The SAM model code is freely available at \url{http://rossby.msrc.sunysb.edu/SAM.html}. The main simulation outputs and related statistics are freely available in \cite{silvestri_2026_24336231}.








%



\appendix[A] 

\appendixtitle{Derivation of generalized Yaglom equation}

The generalized Yaglom equation is the evolution equation for the second-order statistical moment of the two-point increment of frozen moist static energy per unit volume,
\begin{equation}
    \widetilde{\rhobar} \langle \delta h_f'^2 \rangle = \widetilde{\rhobar} \langle \delta h_f' \delta h_f' \rangle
\end{equation}
where 
\begin{equation}
    \delta h_f'   \equiv h_f'(x_i^1,t)-h_f'(x_i^2,t)
\end{equation}
and $\vb{x^1}=x_i^1$ and $\vb{x^2}=x_i^2$ are two generic points in the flow solution. In the above relations, the operators $\delta (\cdot)$ and $\widetilde{(\cdot)}$ denote the two-points increment and average that for a generic quantity $\gamma$ read
\begin{equation}
    \delta \gamma   \equiv \gamma (x_i^1,t)- \gamma (x_i^2,t) \qquad \widetilde{\gamma}   \equiv (\gamma (x_i^1,t) + \gamma (x_i^2,t))/2 \, .
\end{equation}
The theoretical framework has been introduced for the first time by \citet{yaglom1949local} for the scalar temperature field in statistically homogeneous and isotropic turbulence. Here, it is extended to the scalar field of frozen moist static energy and for the more general flow conditions of full statistical inhomogeneity and anisotropy, following the procedure developed by \citet{hill2002exact} for the velocity field.

We first specify the equation for the anomaly of frozen moist static energy (see Appendix A of part I of the present research \cite{silvestri2026gye_part1}),
\begin{align}
    \pdv{h_f'}{t} &+\irhobar \pdv{\rhobar H u'_j}{x_j} + \irhobar \pdv{\rhobar h_f' u'_j}{x_j} + \irhobar \pdv{\rhobar h_f' U_j}{x_j} -\irhobar \pdv{\rhobar  \langle h_f' u'_j \rangle}{x_j}= \nonumber \\
    &=-\irhobar \pdv{(L_f P_f)'}{z}- \irhobar\pdv{\tau_{h_f,j}'}{x_j} + c_{pd}Q_{rad}'
    \label{eq:hp_p1}
\end{align}
where $H = \langle h_f \rangle$ and $U_i = \langle u_i \rangle$ denote the average fields of frozen moist static energy and velocity, $\tau_{h_f,j}$ are the subgrid scalar fluxes, $Q_{rad}$ is the radiative heating rate and $P_f$ are the frozen precipitation fluxes. By writing the above equation for the two points $\vb{x^1}$ and $\vb{x^2}$ and subtracting each other we get,
\begin{align}
    \pdv{h_f'^1}{t} - \pdv{h_f'^2}{t} &+ u_j'^1 \pdv{ H^1}{x_j^1} -u_j'^2 \pdv{H^2}{x_j^2}  + u_j'^1 \pdv{h_f'^1}{x_j^1} - u_j'^2 \pdv{h_f'^2}{x_j^2} + U^1_j \pdv{h_f'^1}{x^1_j} - U^2_j \pdv{h_f'^2}{x^2_j} +\nonumber \\ 
    &-\irhobarp\pdv{\rhobarp  \langle h_f'^1 u_j'^1 \rangle}{x^1_j} + \irhobarpp\pdv{\rhobarpp  \langle h_f'^2 u'^2_j \rangle}{x^2_j}= -\irhobarp \pdv{(L_f P_f)'^1}{z^1} + \irhobarpp \pdv{(L_f P_f)'^2}{z^2} + \nonumber \\
    &- \irhobarp\pdv{\tau_{h_f,j}'^1}{x_j^1} + \irhobarpp\pdv{\tau_{h_f,j}'^2}{x_j^2} + c_{pd} \left (Q_{rad}'^1-Q_{rad}'^2 \right)
    \label{eq:hp_pdiff}
\end{align}
where the superscripts $(\cdot)^1$ and $(\cdot)^2$ has been used to indicate variables computed at the points $\vb{x^1}$ and $\vb{x^2}$, respectively. In the above equation, the mass conservation equation $\pdv*{\rhobar u_i}{x_i}=0$ has been repeatedly used. We can now express the terms involving the anomaly of frozen moist static energy as a function of two-point increments by considering the following relations,
\begin{align}
    \pdv{\delta h_f'}{x_j^1}=\pdv{h_f'^1}{x_j^1}-\cancelto{0}{\pdv{h_f'^2}{x_j^1}}=\pdv{h_f'^1}{x_j^1} \\
    \pdv{\delta h_f'}{x_j^2}=\cancelto{0}{\pdv{h_f'^1}{x_j^2}}-\pdv{h_f'^2}{x_j^2}=-\pdv{h_f'^2}{x_j^2} 
\end{align}
where we have considered that $\vb{x^1}$ and $\vb{x^2}$ are independent variables, so that $\partial{h_f'^1} /\partial {x_j^2}=0$ and $\partial{h_f'^2} /\partial {x_j^1}=0$. By using the above relations and similar ones for the other terms, equation \ref{eq:hp_pdiff} can be rewritten as
\begin{align}
    \pdv{\delta h_f'}{t} &+ u'^1_j \pdv{\delta H}{x_j^1} + u'^2_j \pdv{\delta H}{x_j^2}  + u'^1_j \pdv{\delta h_f'}{x_j^1} + u'^2_j \pdv{\delta h_f'}{x_j^2} + U^1_j \pdv{\delta h_f'}{x^1_j} + U^2_j \pdv{\delta h_f'}{x^2_j} +\nonumber \\ 
    &-\irhobarp\pdv{\delta \big(\rhobar  \langle h_f' u_j' \rangle \big )}{x^1_j} - \irhobarpp\pdv{\delta \big (\rhobar  \langle h_f' u_j' \rangle \big )}{x^2_j} = -\irhobarp \pdv{\delta (L_f P_f)'}{z^1} - \irhobarpp \pdv{\delta (L_f P_f)'}{z^2} + \nonumber \\
    &- \irhobarp\pdv{\delta \tau_{h_f,j}'}{x_j^1} - \irhobarpp\pdv{\delta \tau_{h_f,j}'}{x_j^2} + c_{pd}\delta Q_{rad}'
\end{align}
We can now introduce the new set of independent variables $\vb{r}=\vb{x^1}-\vb{x^2}$ and $\vb{x_c}=(\vb{x^1}+\vb{x^2})/2$. In particular, by considering the following differential relations,
\begin{align}
    \pdv{x_i^1}=\frac{1}{2}\pdv{x_{c_i}}+\pdv{r_i} \\
    \pdv{x_i^2}=\frac{1}{2}\pdv{x_{c_i}}-\pdv{r_i}
\end{align}
it is possible to obtain the final form of the equation for the two-point increments of the anomaly of frozen moist static energy $\delta h_f'$,
\begin{align}
    \pdv{\delta h_f'}{t} &+ \widetilde{u_j} \pdv{\delta H}{x_{c_j}} + \delta u_j' \pdv{\delta H}{r_j}  + \widetilde{u_j} \pdv{\delta h_f'}{x_{c_j}} + \delta u_j' \pdv{\delta h_f'}{r_j} + \widetilde{U_j} \pdv{\delta h_f'}{x_{c_j}} + \delta U_j \pdv{\delta h_f'}{r_j} +\nonumber \\ 
    &-\frac{1}{2}\left(\irhobarp + \irhobarpp \right)\pdv{\delta(\rhobar  \langle h_f' u_j' \rangle)}{x_{c_j}} -\left(\irhobarp - \irhobarpp \right) \pdv{\delta(\rhobar  \langle h_f' u_j' \rangle)}{r_j} = \nonumber \\
    &=-\frac{L_f}{2} \left( \irhobarp + \irhobarpp \right ) \pdv{\delta P_f'}{z_{c}} -L_f \left (\irhobarp - \irhobarpp \right ) \pdv{\delta P_f'}{r_j} + \nonumber \\
    &- \frac{1}{2} \left ( \irhobarp + \irhobarpp \right ) \pdv{\delta \tau_{h_f,j}'}{x_{c_j}} - \left ( \irhobarp - \irhobarpp \right ) \pdv{\delta \tau_{h_f,j}'}{r_j} + c_{pd}\delta Q_{rad}'
\end{align}
In order to obtain the equation for the second-order moment of the energy increment, we can now multiply the equation by $2\delta h_f'$ and we can average. The result is the Yaglom equation generalized for fully anisotropic and inhomogeneous flows,
\begin{align}
    \pdv{ \langle \delta h_f'^2 \rangle}{t} &+ 2 \langle \delta h_f' \widetilde{u_j'} \rangle \, \delta \left (\pdv{H}{x_j} \right ) + 2 \langle \delta h_f' \delta u_j' \rangle \widetilde{\left (\pdv{H}{x_j} \right )}  +  \langle \widetilde{u_j'} \pdv{\delta h_f'^2}{x_{c_j}} \rangle +  \langle \delta u_j' \pdv{\delta h_f'^2}{r_j} \rangle + \nonumber \\
    &+  \widetilde{U_j} \pdv{\langle \delta h_f'^2 \rangle}{x_{c_j}} + \delta U_j \pdv{\langle \delta h_f'^2 \rangle}{r_j}  =\nonumber \\ 
    &=- L_f \left(\irhobarp + \irhobarpp \right) \langle \delta h_f' \pdv{\delta P_f'}{z_{c}} \rangle - 2 L_f \left (\irhobarp - \irhobarpp \right ) \langle \delta h_f' \pdv{\delta P_f'}{r_j} \rangle + \nonumber \\
    &- \left ( \irhobarp + \irhobarpp \right ) \langle \delta h_f' \pdv{\delta \tau_{h_f,j}'}{x_{c_j}} \rangle - 2 \left ( \irhobarp - \irhobarpp \right ) \langle \delta h_f' \pdv{\delta \tau_{h_f,j}'}{r_j} \rangle + 2 c_{pd} \langle \delta h_f' \delta Q_{rad}' \rangle
    \label{eq:full_gye_app}
\end{align}
where we have considered that $\partial \delta H / \partial x_{c_j} = \delta (\partial H / \partial x_j)$ and $\partial \delta H / \partial r_j = \widetilde{(\partial H / \partial x_j)}$. The generalized Yaglom equation (\ref{eq:full_gye_app}) describes the evolution of the second-order structure function of frozen moist static energy in the full seven-dimensional augmented space, composed by the three-dimensional space of positions $\vb{x_c}$, the three-dimensional space of scales $\vb{r}$ and time $t$, $\langle \delta h_f'^2 \rangle = \langle \delta h_f'^2 \rangle (\vb{r}, \vb{x_c}, t)$.

When considering the statistical symmetries of the of the radiative-convective equilibrium problem, the dimensionality and complexity of the problem described by the generalized Yaglom equation (\ref{eq:full_gye_app}) drastically reduces. In particular, the statistical steadiness, the homogeneity and isotropy in the horizontal planes and the absence of a mean wind simplify the generalized Yaglom equation as,
\begin{align}
    2 \langle \delta h_f' \widetilde{w'} \rangle \, \delta \left (\pdv{H}{z} \right ) &+ 2 \langle \delta h_f' \delta w' \rangle \widetilde{\left (\pdv{H}{z} \right )}  +  \langle \widetilde{u_j'} \pdv{\delta h_f'^2}{x_{c_j}} \rangle +  \langle \delta u_j' \pdv{\delta h_f'^2}{r_j} \rangle = \nonumber \\
    &=- L_f \left(\irhobarp + \irhobarpp \right) \langle \delta h_f' \pdv{\delta P_f'}{z_{c}} \rangle - 2 L_f \left (\irhobarp - \irhobarpp \right ) \langle \delta h_f' \pdv{\delta P_f'}{r_j} \rangle + \nonumber \\
    &- \left ( \irhobarp + \irhobarpp \right ) \langle \delta h_f' \pdv{\delta \tau_{h_f,j}'}{x_{c_j}} \rangle - 2 \left ( \irhobarp - \irhobarpp \right ) \langle \delta h_f' \pdv{\delta \tau_{h_f,j}'}{r_j} \rangle + 2 c_{pd} \langle \delta h_f' \delta Q_{rad}' \rangle
\end{align}
and the augmented space reduces to a three dimensional space, $\langle \delta h_f'^2 \rangle = \langle \delta h_f'^2 \rangle (\sqrt{r_x^2+r_y^2}, r_z, z_c)$, composed by a two-dimensional space of scales $(\sqrt{r_x^2+r_y^2}, r_z)$ and by a one-dimensional space of atmospheric elevations $z_c$.

The problem can be further simplified by limiting the analysis to the plane $r_z=0$, as we do in the present work. In this space, $\rhobar (z_c) = \rhobarp = \rhobarpp$, $\delta (\partial H / \partial z) = 0$ and $(\partial H / \partial z) (z_c) = \widetilde{(\partial H / \partial z)}$. Hence, the generalized Yaglom equation reduces to
\begin{align}
    2 \langle \delta h_f' \delta w' \rangle \left (\pdv{H}{z} \right )  &+  \langle \widetilde{u_j'} \pdv{\delta h_f'^2}{x_{c_j}} \rangle +  \langle \delta u_j' \pdv{\delta h_f'^2}{r_j} \rangle =  \nonumber \\
    &= - \frac{2 L_f}{\rhobar} \langle \delta h_f' \, \delta \left (\pdv{P_f'}{z} \right ) \rangle - \frac{2}{\rhobar} \langle \delta h_f' \pdv{\delta \tau_{h_f,j}'}{x_{c_j}} \rangle + 2 c_{pd} \langle \delta h_f' \delta Q_{rad}' \rangle
    \label{eq:gye_app}
\end{align}
To easy the physical interpretation of the equation, it is now possible to write it in a conservative form by rewriting different terms in a divergence form. Starting from the diffusive term, after some arrangement it can be first rewritten as
\begin{equation}
    - \frac{2}{\rhobar} \langle \delta h_f' \pdv{\delta \tau_{h_f,j}'}{x_{c_j}} \rangle = - \frac{1}{\rhobar} \langle \delta h_f' \pdv{\delta \tau_{h_f,j}'}{x_{c_j}} \rangle-\frac{4}{\rhobar} \langle \delta h_f' \pdv{\widetilde{\tau_{h_f,j}'}}{r_j} \rangle
\end{equation}
The right hand side can be further expanded by considering the following relations
\begin{align}
    - \frac{1}{\rhobar} \langle \delta h_f' \pdv{\delta \tau_{h_f,j}'}{x_{c_j}} \rangle&=- \frac{1}{\rhobar} \pdv{\langle \delta \tau_{h_f,z}' \delta h_f' \rangle}{z_{c}} + \irhobar \langle \delta \tau_{h_f,j}'\pdv{\delta h_f'}{x_{c_j}} \rangle= \nonumber \\
    &=- \frac{1}{\rhobar} \pdv{\langle \delta \tau_{h_f,z}' \delta h_f' \rangle}{z_c} + \irhobar \langle (\tau_{h_f,j}'^1-\tau_{h_f,j}'^2)\left (\pdv{h_f'^1}{x_j^1}-\pdv{h_f'^2}{x_j^2}\right ) \rangle= \nonumber\\
    &=- \frac{1}{\rhobar} \pdv{\langle \delta \tau_{h_f,z}' \delta h_f' \rangle}{z_c} + \irhobar\left ( \langle \tau_{h_f,j}'^1\pdv{h_f'^1}{x_j^1} \rangle+ \langle \tau_{h_f,j}'^2\pdv{h_f'^2}{x_j^2} \rangle - \langle \tau_{h_f,j}'^1\pdv{h_f'^2}{x_j^2} \rangle - \langle \tau_{h_f,j}'^2\pdv{h_f'^1}{x_j^2} \rangle \right ) \nonumber \\
    - \frac{4}{\rhobar} \langle \delta h_f' \pdv{\widetilde{\tau_{h_f,j}'}}{r_j} \rangle&=- \frac{4}{\rhobar} \pdv{\langle \widetilde{\tau_{h_f,j}'} \delta h_f' \rangle}{r_j} + \frac{4}{\rhobar} \langle \widetilde{\tau_{h_f,j}'}\pdv{\delta h_f'}{r_j} \rangle= \nonumber \\
    &=- \frac{4}{\rhobar} \pdv{\langle \widetilde{\tau_{h_f,j}'} \delta h_f' \rangle}{r_j} +\frac{1}{\rhobar} \langle (\tau_{h_f,j}'^1+\tau_{h_f,j}'^2) \left (\pdv{h_f'^1}{x_j^1}+\pdv{h_f'^2}{x_j^2} \rangle \right)= \nonumber \\
    &=- \frac{4}{\rhobar} \pdv{\langle \widetilde{\tau_{h_f,j}'} \delta h_f' \rangle}{r_j} +\frac{1}{\rhobar} \left( \langle \tau_{h_f,j}'^1\pdv{h_f'^1}{x_j^1} \rangle+ \langle \tau_{h_f,j}'^2\pdv{h_f'^2}{x_j^2} \rangle + \langle \tau_{h_f,j}'^1\pdv{h_f'^2}{x_j^2} \rangle + \langle \tau_{h_f,j}'^2\pdv{h_f'^1}{x_j^2} \rangle \right) \nonumber 
\end{align}
thus allowing us to rewrite the diffusion term as
\begin{equation}
    - \frac{2}{\rhobar} \langle \delta h_f' \pdv{\delta \tau_{h_f,j}'}{x_{c_j}} \rangle = - \frac{1}{\rhobar} \pdv{\langle \delta \tau_{h_f,z}' \delta h_f' \rangle}{z_c} - \frac{4}{\rhobar} \pdv{\langle \widetilde{\tau_{h_f,j}'} \delta h_f' \rangle}{r_j} - \frac{4}{\rhobar} \langle \epsilon_{sgs}^h \rangle
    \label{eq:diss_app}
\end{equation}
where $\epsilon_{sgs}^h=- \tau_{h_f,j}'(\partial h_f'/\partial x_j)$. To further highlight the conservative form of the generalized Yaglom equation, it is possible to write also the advective terms in a divergence form by multiplying the equation by $\rhobar$, 
\begin{align}
    &\rhobar \langle \widetilde{u_j'} \pdv{\delta h_f'^2}{x_{c_j}} \rangle = \pdv{\rhobar\langle \delta h_f'^2 \widetilde{w'} \rangle}{z_c}   \nonumber \\
    &\rhobar \langle \delta u_j' \pdv{\delta h_f'^2}{r_j} \rangle = \pdv{\rhobar  \langle \delta h_f'^2 \delta u_j' \rangle}{r_j}
    \label{eq:adv_app}
\end{align}
where the mass conservation conditions $\partial \rhobar \widetilde{u_j'} / \partial x_{c_j} = 0$ and $\partial \rhobar \delta u_j' / \partial r_j = 0$ have been used. Overall, by multiplying equation (\ref{eq:gye_app}) by $\rhobar$ and by substituting relations (\ref{eq:diss_app}) and (\ref{eq:adv_app}), the final form of the generalized Yaglom equation for the statistical symmetries of the radiative-convective equilibrium problem and for the $(r_z=0)$-space, is
\begin{align}
    &\pdv{\rhobar  \langle \delta h_f'^2 \delta u_j' \rangle}{r_j}+ \pdv{\rhobar\langle \delta h_f'^2 \widetilde{w'} \rangle}{z_c} + \pdv{\langle \delta \tau_{h_f,z}' \delta h_f' \rangle}{z_c} + 4 \pdv{\langle \widetilde{\tau_{h_f,j}'} \delta h_f' \rangle}{r_j} = \nonumber \\
    &= - 2 \rhobar \langle \delta h_f' \delta w' \rangle \left (\pdv{H}{z} \right ) - 4 \langle \epsilon_{sgs}^h \rangle  - 2 L_f \langle \delta h_f' \, \delta \left (\pdv{P_f'}{z} \right ) \rangle + 2 \rhobar c_{pd} \langle \delta h_f' \delta Q_{rad}' \rangle
\end{align}


\bibliographystyle{ametsocV6}
\bibliography{references}

\end{document}